\documentclass[usenatbib,]{mnras}

\usepackage{newtxtext,newtxmath}
\usepackage[T1]{fontenc}
\usepackage{ae,aecompl}

\usepackage[utf8]{inputenc}
\usepackage{textcomp}
\usepackage{gensymb}
\usepackage{amsmath}

\usepackage{amssymb}
\usepackage{xspace}
\usepackage{nicefrac}
\usepackage{xcolor}
\usepackage{graphicx}
\usepackage{soul}
\usepackage{hyperref}
\usepackage[caption=false]{subfig}
\usepackage{multirow}
\usepackage{cleveref}
\usepackage{enumitem}
\usepackage[titletoc,toc,page]{appendix}
\usepackage{xcolor}
\usepackage{rotating}

\usepackage{threeparttable}

\DeclareMathAlphabet\mathrsfso      {U}{rsfso}{m}{n}

\usepackage{esint}

\begin{document}

\title{The origin of the very-high-energy radiation along the jet of Centaurus A}

%\author{Cain\~a de Oliveira}
%\email{caina.oliveira@usp.br}
%\affiliation{Instituto de F\'isica de S\~ao Carlos, Universidade de S\~ao Paulo, Av. Trabalhador S\~ao-carlense 400, S\~ao Carlos, Brasil.}
%\author{James Matthews}
%\email{james.matthews@physics.ox.ac.uk}
%\affiliation{Department of Physics, Astrophysics, University of Oxford, Denys Wilkinson Building, Keble Road, Oxford, OX1 3RH, UK}
%\author{Vitor de Souza}
%\email{vitor@ifsc.usp.br}
%\affiliation{Instituto de F\'isica de S\~ao Carlos, Universidade de S\~ao Paulo, Av. Trabalhador S\~ao-carlense 400, S\~ao Carlos, Brasil.}

\author[C. de Oliveira et al.]{
Cainã de Oliveira,$^{1}$\thanks{caina.oliveira@usp.br; olivcaina@gmail.com}
James H. Matthews,$^{2}$
and Vitor de Souza$^{1}$
\\
$^{1}$Instituto de F\'isica de S\~ao Carlos, Universidade de S\~ao Paulo, Av. Trabalhador S\~ao-carlense 400, S\~ao Carlos, Brasil.\\
$^{2}$Department of Physics, Astrophysics, University of Oxford, Denys Wilkinson Building, Keble Road, Oxford, OX1 3RH, UK}

%\date{\today}

\maketitle

\begin{abstract}   
    As the closest known active galactic nucleus, Centaurus A (Cen~A) provides a rich environment for astrophysical exploration. It has been observed across wavelengths from radio to gamma rays, and indications of ongoing particle acceleration have been found on different scales. Recent measurements of very-high-energy (VHE) gamma-rays ($>240$~GeV) by the HESS observatory have inferred the presence of ultra-relativistic electrons along Cen A’s jet, yet the underlying acceleration mechanism remains uncertain. Various authors have proposed that jet substructures, known as knots, may serve as efficient particle accelerators. In this study, we investigate the hypothesis that knots are the particle acceleration sites along Cen A’s jets. We focus on stationary knots, and assume that they result from interactions between the jet and the stellar winds of powerful stars. By combining relativistic hydrodynamic simulations and shock acceleration theory with the radio and X-ray data, we compare theoretical predictions with morphological and spectral data from different knots. We estimate the maximum electron energy and the resulting VHE gamma-ray emission. Our findings suggest that electrons accelerated at the knots are responsible for the gamma-ray spectrum detected in the VHE band.
\end{abstract}

\begin{keywords}
galaxies: jets -- hydrodynamics -- shock waves -- acceleration of particles --  quasars: individual: Centaurus A -- gamma-rays: galaxies
\end{keywords}

\section{Introduction}
%Gamma ray in AGNs

%Centaurus A
Centaurus A (NGC~5128, Cen~A) is the closest active galactic nucleus (AGN), at a distance of $3.8 \pm 0.1 $~Mpc from the Milky Way~\citep{harris2010distance}. It produces collimated jets on kiloparsec scales and ejects galactic material at scales of $\sim0.5$~Mpc forming large-scale lobes~\citep{israel1998centaurus,mckinley2022multi}.

Cen~A has been studied at wavelengths from radio to $\gamma$-rays~\citep{israel1998centaurus}. Concerning radio morphology, it is classified as an FRI radio galaxy, with a radio luminosity $2.6 \times 10^{40}$~erg~s$^{-1}$ measured at $1.40$~GHz~\citep{van2012radio}. The proximity of Cen~A allows substructures to be resolved in radio and X-ray both in the lobes and jet~\citep{mckinley2022multi, Bogensberger_2024,prabu2024subparsec}, revealing the presence of nonthermal radiation as diffuse emission and in the form of knots~\citep{Kataoka_2006,Worrall_2008,Goodger_2010,Snios_2019}. 

Anisotropy studies conducted by the Pierre Auger Collaboration reveal that Cen~A is located in the direction of an excess of UHECR events~\citep{caccianiga2021anisotropies}. Being a nearby AGN, there are several works pointing to Cen~A being the UHECR source causing the excess ~\citep{ginzburg1963cosmic,ROMERO1996279,liu2012excess,Farrar_2013,wykes_2013,wykes2018uhecr,joshi2018very,bib:biermann:1}. It is known that nearby sources are necessary to explain the more energetic parcel of UHECR~\citep{lang2020revisiting}, however, the sources of UHECRs were never detected definitively, remaining one of the largest puzzles of current astrophysics~\citep{alves2019open}. UHECRs from distant sources are subject to higher deflections in extragalactic magnetic fields~\citep{achterberg1999intergalacticpropagationuhecosmic}, making Cen~A one of the most compelling detectable candidates for UHECR accelerators due to its proximity and power.

Cen~A is an important source of high-energy radiation, being observed by different instruments~\citep{Hartman_1999,fermi_science_2010,fermi_hess_2018,hess2020resolving}. The central source~\citep{fermi_hess_2018}, the giant lobes~\citep{fermi_science_2010}, and potentially the inner lobes~\citep{fermi_inner_cenA}, are $\gamma$-ray sources. Between GeV and TeV energies, the spectrum of the central source suffers an upturn that prevents it from being described by one single zone Synchrotron Self-Compton model for the core, requiring an additional component~\citep{fermi_hess_2018}.
Different explanations for the origin of the VHE radiation observed in Cen~A have been proposed (see, for example, Figure 4 of \citep{Rieger_2017}). The recent discovery of extended $\gamma$-ray emission by \citet{hess2020resolving}, was interpreted as evidence for ultra-relativistic electrons in the jet. However, the acceleration mechanism remains unknown.

\citet{Tanada_2019} and \citet{Sudoh_2020} proposed the knots detected along the jet of Cen~A as the main candidates for particle (re)acceleration. Knots are clumps of bright material that can originate from turbulence, fluid compression, collisions with obstacles, jet re-confinement, or magnetic reconnection. The presence of knots in the jet of Cen~A is ubiquitous, being detected on parsec and kiloparsec scales, both on radio and X-ray wavelengths~\citep{Goodger_2010,tanami2014}. Most of the Cen~A's knots can be explained by collisions of the jet with obstacles, such as stellar winds (SW), planetary nebulae, or gas/molecular clouds~\citep{Hardcastle_2003,Goodger_2010,Snios_2019}. The collision produces shock complexes and regions of magnetic field amplification suitable for particle acceleration~\citep{bednared_1997,araudo_2013,Cita_2016}. Two newly identified X-ray knots, C1 and C2~\citep{Bogensberger_2024}, could be attributed to impulsive particle acceleration (based only on the timescale of the bright variation), but it is unclear if they belong to the jet.

First order Fermi acceleration at shocks (Fermi-I) is the main process proposed to accelerate particles in jet-SW interactions~\citep{bednared_1997,araudo_2013}. The collision generates a double shock structure separated by a contact discontinuity. Cosmic rays and electrons can be accelerated in wind and jet shocks. The interaction also causes turbulence, which mixes the jet material with the wind and can cause the deceleration of the jet and internal entrainment of mass~\citep{wykes_2013,wykes_2015,wykes1d}.

The accelerated electrons will produce non-thermal radiation mainly due to synchrotron and Inverse Compton (IC) losses. Synchrotron radiation can explain the radio and X-ray emission detected~\citep{Goodger_2010,Snios_2019}. The resultant $\gamma$-ray emission from IC should have energies up to $\sim100$~TeV, with peaks in the range $0.01-1$~TeV~\citep{araudo_2013, Cita_2016}, and has been proposed to explain the spectrum and variability from GeV to TeV in astrophysical jets~\citep{bednared_1997,Barkov_2012,araudo_2013,Cita_2016,Cita_2017}.

%--------------------%

In this work, we conduct a detailed analysis of the hypothesis of knots generated by the collision of the jet with a SW in Cen~A. We combine relativistic hydrodynamical (RHD) simulations with a radiation model to solve the knot structure and predict the VHE $\gamma$ radiation from the knots and their immediate surroundings. For the first time, we combine morphology and multi-wavelength measurements of Cen~A with fluid simulations and particle acceleration theory to test the hypothesis of jet-SW collision as the origin of knots and high-energy radiation in Cen~A's jet. In section~\ref{sec:model} the knot and acceleration models are described. The results for the brighter AX1A, AX1C, AX2, and BX2 knots are presented in section~\ref{sec:model_results}. In section~\ref{sec:extended} the expected emission of the extended region around the knots and from knots complexes is discusses. The implications of the model are presented in section~\ref{sec:implications}. Finally, section~\ref{sec:conclusions} summarizes the main conclusions found in this work.

\begin{figure*}
    \centering
    \includegraphics[width=\textwidth]{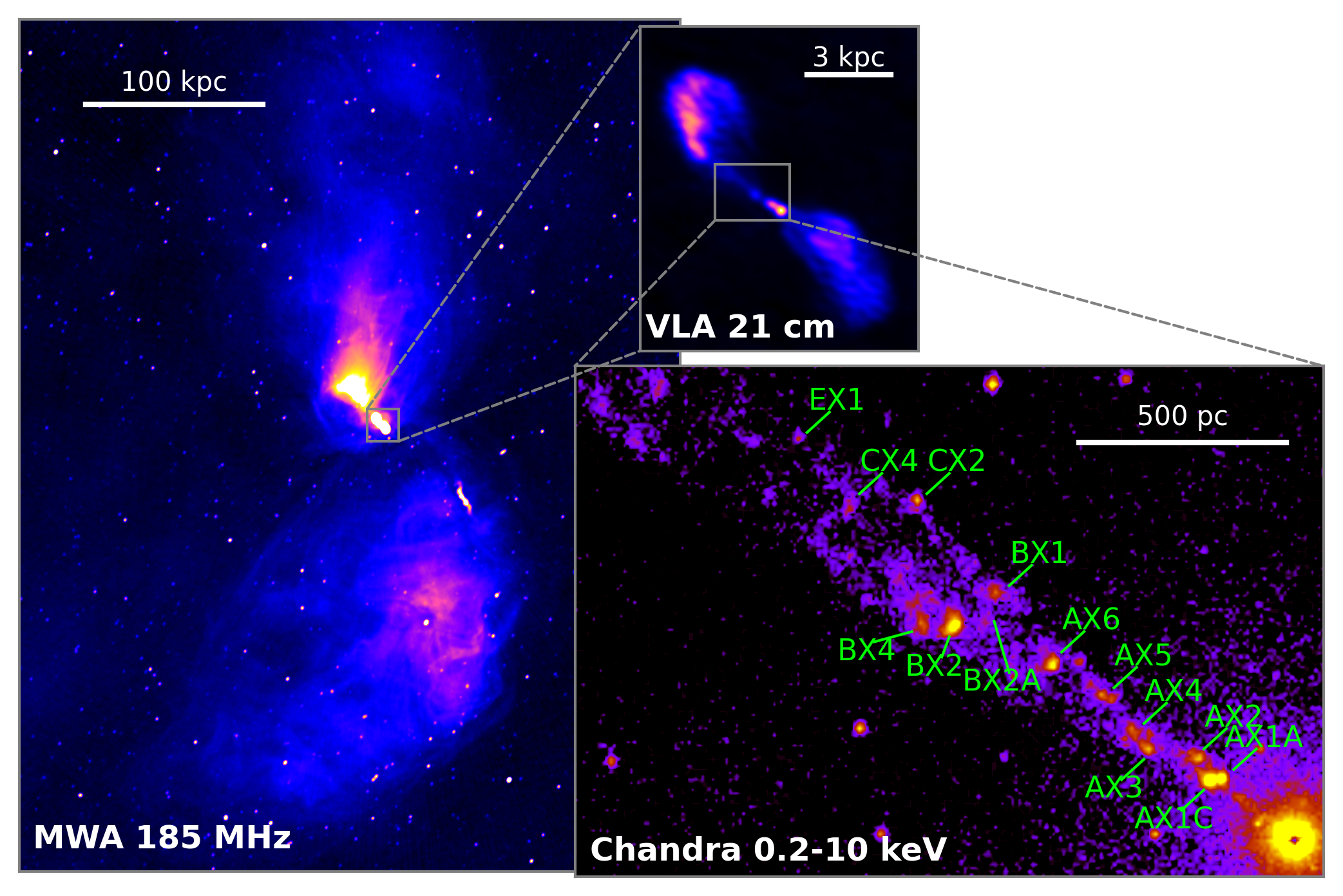}
    \caption{Radio and X-ray images of Cen~A. {\sl Left:} Murchison Widefield Array (MWA) radio (185 MHz) image of the whole radio source~(\citealt{mckinley2022multi}; \url{https://cdsarc.cds.unistra.fr/viz-bin/cat/J/other/NatAs/6.109}, colors in logarithm scale). {\sl Top right:} Very Large Array (VLA) radio (21 cm or 1.4~GHz) image of the kiloparsec jet and inner lobes~\citep[][colors in linear scale]{condon19961}. {\sl Bottom right:} X-ray (0.2--10 keV) image of the jet as seen by \textit{Chandra} (ObsID=20794, \url{https://doi.org/10.25574/20794}, PI: P. Nulsen, see also ~\citealt{Snios_2019}, colors in logarithm scale). The main knots are labeled following \citet{Snios_2019}. A bilinear interpolation smoothing was applied to all images.}
    \label{fig:cenA}
\end{figure*}

\section{Modeling the knots}
\label{sec:model}
Cen A's knots have been detected both in radio and X-ray wavelengths, by VLA and \textit{Chandra}, respectively~\cite [e.g.][]{Hardcastle_2003,Goodger_2010, Snios_2019}. An image showing the X-ray jet and knots of Cen~A is presented in figure~\ref{fig:cenA}, alongside the radio observation of the giant lobes and the kiloparsec jet with the inner lobe. To develop a consistent model, we focus on the brightest, better-resolved, AX1A, AX1C, AX2, and BX2 knots, following \citet{Tanada_2019}. These knots are stationary considering 22 years of X-ray observations~\citep{Bogensberger_2024}, consistent with radio measurements~\citep{Goodger_2010}. VLBI radio images~\citep{Tingay_2009} support the collision model for the origin of the stationary knots with X-ray counterparts by detecting the compact centroid of knots A1A, A1C, and A2A, radio counterparts of AX1A, AX1C, and AX2, respectively (BX2 was not present in the analysis). The knot A1A, and possibly A2A, present substructures in the VLBI images. The main properties of the X-ray knots used in this work can be found in Table~\ref{tab:knots_data}. AX1A and AX1C could also be associated with a re-confinement shock, but the fact that the knots present substructure and the presence of the A1B knot moving between them challenge this hypothesis~\citep{Goodger_2010}.

Our approach to modelling the knots can be separated into the following steps:
\begin{enumerate}[label=\textbf{\Alph*.}]
    \item Solving the fluid equations for the jet's interaction with a stellar wind (section~\ref{sec:fluid}). The jet properties are taken from data, and the stellar wind is assumed based on that of proposed stars generating the knots;
    \item Determination of the electron population (section~\ref{sec:elec}). Nonthermal electrons are assumed to be accelerated at the shock waves produced in the jet/stellar wind interaction and evolve through the computational grid. Synchrotron and Inverse Compton cooling are also taken into account;
    \item Prediction of nonthermal radiation from the electrons (section~\ref{sec:sed}). The model's free parameters are obtained based on the radio and X-ray measurements for the knots, and the very-high-energy $\gamma$ rays are predicted.
\end{enumerate}
Below each step is described in detail.

\begin{table*}
    \centering
    \begin{threeparttable}
    \caption{Experimental estimations for the X-ray knots AX1A, AX1C, AX2 and BX2 and its radio counter-part A1A (a/b), A1C, A2A (a/b) and B1A.}
    \begin{tabular}{c|c|c|c|c|c}
        Knot & $z_{\rm knot}$ (pc)\tnote{a} & $R_{\rm j}$ (pc)\tnote{b} & $\sigma_{\rm VLBI}$ (pc)\tnote{c} & $\sigma_{\rm X}$ (pc)\tnote{d} & $\sigma_{\rm X}$/ $R_{\rm j}$\\
        \hline
        AX1A / A1A (a/b) & 292 & 33 & 1.65 / 2.1 & 2.6 & 0.08 \\
        AX1C / A1C & 292 & 33 & 4.8 & 5.5 & 0.17 \\
        AX2 / A2A (a/b) & 356 & 37 & 2.7 / 5.9 & 8.8 & 0.24 \\
        BX2 / B1A & 1134 & 63 & -  & 11.3 & 0.18 \\ 
    \end{tabular}
    \label{tab:knots_data}
    \begin{tablenotes}
        \item[a] Distance of the knot to Cen~A's core, obtained by \citet{Kraft_2002}.
        \item[b] Jet radius at $z=z_{\rm knot}$ evaluated using the measurements from \textit{Chandra} of~\cite{wykes1d}.
        \item[c] Diameter of the radio counterpart estimated based on the VLBI measurements of \citet{Tingay_2009}. The VLBI meaurements indicates substructures for A1A and A2A, labelled as \textit{a} and \textit{b}.
        \item[d] X-ray knot full width at maximum height (FWMH), based on \textit{Chandra} measurements obtained by \citet{Tanada_2019}.
    \end{tablenotes}
    \end{threeparttable}
\end{table*}

\subsection{Fluid}\label{sec:fluid} %see https://www.arxiv.org/pdf/2409.05256
To develop an underlying model for the interaction between the jet of Cen~A and the stellar wind, the initial conditions of the RHD simulation were informed by the available data from radio and X-ray observations of Cen~A. The RHD equations were solved in an axisymmetric 2D cylindrical coordinate frame using the {\fontfamily{cmtt}\selectfont PLUTO} code~\citep{Mignone_2007}. The computational grid is measured in units of the jet radius ($R_j$), with $r_{\rm sim} = r / R_j$, $z_{\rm sim} = z / R_j$, with $z$ along the jet axis and $r$ as the cylindrical radius. The value of $R_j$ is taken at the distance between $z_{\rm knot}$ and the core. Variations of the radius of the jet in the definition of the grid are neglected. The simulation was performed in the interval $r_{\rm sim} = [0, 1.5]$, $z_{\rm sim} = [0, 4.0]$, with resolution $\delta_r = \delta_z = 1/200$ leading to a grid of $300 \times 800$ cells. Around the BX2 knot, for instance, the simulation grid encompasses $z=1102$~pc up to $z=1354$~pc, and considering $R_j = 63$~pc, the change in the jet radius between the base and the top of the grid is $\sim 6$~pc, a difference $< 10\%$.

The spatial integration order follows a piecewise TVD linear reconstruction (second-order accurate), with a harmonic mean limiter of van Leer. The Riemann problem was solved using the Harten, Lax, Van Leer (HLLC) approximation \citep{mignone_hllc}. Additional dissipation was applied using the standard multi-dimensional shock flattening strategy to increase the code robustness. The time increment was evaluated using a second-order TVD Runge Kutta algorithm with a Courant-Friedrichs-Levy number of 0.4.

We do not consider the interaction of the jet with the external medium in the present analysis. The initial condition of the simulation is set by the jet properties, except for in cells within the stellar wind injection region.

\subsubsection{Jet}
%Velocity
Based on the X-ray data provided by \textit{Chandra} measurements, \citet{Snios_2019} found an apparent speed for the collective proper motion of the fainter substructure in the jet of $(0.68 \pm 0.20)c$ (equivalent to $(0.57 \pm 0.11)c$, for a jet inclination angle of $50^\circ$), over the projected length of $0.26-1.35$~kpc along the jet of Cen~A. The speed is comparable to the values from VLA's radio data~\citep{Goodger_2010}. Considering individual knots, \citet{Bogensberger_2024} reported an apparent speed of $(0.46 \pm 0.11)c$, where the differences are attributed to bright knots and diffuse material. A superluminal apparent velocity was detected for the knot AX4, which requires a minimum velocity of $(0.94 \pm 0.02)c$ for this particular knot. Radio observations reveal a complex, non-laminar motion in the region of the A-knots~\citep{Goodger_2010}. An apparent perpendicular velocity to the jet axis is also detected in X-ray~\citep{Bogensberger_2024}. The turbulence could be associated with the formation of the knots by the interaction of the jet with stellar winds~\citep{wykes1d}. In this work, the velocity of the plasma in the jet will be assumed to be $v_j = 0.6c$ along the jet axis ($\hat{z}$), with no $\hat{r}$ component.

%Pressure
\citet{wykes1d} uses X-ray measurements of \textit{Chandra} to determine the pressure profile of the interstellar medium in the vicinity of the jet, given by
\begin{equation}
    P(r,z) = P_0 \left( \frac{\sqrt{r^2 + z^2}}{\text{kpc}} \right)^{-1.5},
\end{equation}
where $P_0=5.7 \times 10^{-11}$~dyn~cm$^{-2}$ and $\sqrt{r^2 + z^2}$ is the distance from Cen~A's core. Following the work of \citet{wykes1d}, pressure equilibrium between the jet and the external medium is assumed.
A Taub-Matthews equation of state is considered for the plasma in the jet, with $\gamma_{\rm ad}^{\rm jet} = 13/9$, suitable for temperatures where the electrons are relativistic and the protons are non-relativistic~\citep{wykes1d}.

%Density
The jet power of Cen~A has been determined by different methods, being consistently estimated as $L_j \sim10^{43}-10^{44}$~erg~s$^{-1}$~\citep{croston_2009,yang_2012,wykes_2013,Neff_2015,sun_2016}. In this work, $L_j = 2 \times 10^{43}$~erg~s$^{-1}$ will be used. The total power in the jet is given by the sum of the kinetic, thermal, and magnetic contributions,
\begin{equation}
    L_j = \pi R_j^2 v_j \Bigg[ \Gamma_j(\Gamma_j - 1)\rho_j c^2 + \frac{\gamma_{\rm ad}}{\gamma_{\rm ad}-1} \Gamma_j^2 \Big(P_j + B^2/8\pi \Big)\Bigg],
\end{equation}
where $\Gamma_j = \big(1-(\nicefrac{v_j}{c})^2 \big)^{-1/2}$ is the jet bulk Lorentz factor, $P_j$ is the thermal pressure in the jet, and $B$ is the magnetic field intensity. The contribution of the thermal pressure to the total power is $\sim5\%$, indicating that the jet of Cen~A is kinetically dominated, so we neglect the thermal pressure in our work. The magnetic pressure must be relatively unimportant, since the upper limit of the magnetic field in the knot, $\sim80~\mu$G~\citep{Snios_2019}, provides a pressure of the same order as the thermal pressure, and this magnetic field in the knot is likely to be amplified. Knowing $v_j$ and $R_j$, and considering the power of the jet constant in time, the jet density ($\rho_j$) can be estimated. Random perturbations $\delta \rho_j / \rho_j = 10^{-2}$ were applied to break the homogeneity of the jet's fluid. The profile of jet radius along the jet axis, $R_j(z)$, was determined by~\citet{wykes1d} using X-ray measurements. Taking the values at the position of the AX1A/AX1C knots, gives $\rho_j \sim 7 \times 10^{-5}$~cm$^{-3}$.

\subsubsection{Star}

The Stellar Wind (SW) is characterized by the wind terminal velocity ($V_\infty$) and the mass injection rate ($\dot{M}$), both of which depending on the star's class. The parsec scale of the knots' size demands the interaction of the jet with stars in enhanced mass-loss stages. \citet{Hardcastle_2003} propose that rarer Wolf-Rayet (WR) stars may generate knots with radius $\sim 10$~pc. VLBI observations \citet{Tingay_2009} indicates the knot radius $\sim0.5-3$~pc and \citet{Goodger_2010} argues the more common O/B stars as possible obstacles. \citet{wykes_2015,wykes1d} explores the mass entrainment of material from Asymptotic Giant Branch (AGB) stars as responsible for the jet deceleration and proposes that the collision with their SW may be quantitatively responsible for the X-ray emission of the jet in the downstream region ($\sim3$~pc from the galactic center). Luminous Blue Variable stars were also considered~\citep{wykes_2015}. \citet{Muller_2023_conf} have shown that at least one WR star will be found inside the first $\sim200$~pc of AGN jets. This work will explore WR-like stars since they present the most powerful winds, assuming $V_\infty = 3000$~km~s$^{-1}$, and $\dot{M} = 10^{-4}~M_{\odot}$~yr$^{-1}$~\citep{WR_WN,Muller_2023_conf}.

The SW is modeled as a constant flow, radially blown from the knot position $z \approx z_{\rm knot}$. In the reference frame used in the simulation, for each knot, the SW position is chosen as ($r_{\rm sim} = 0, z_{\rm sim}=0.5$). Inside the inner boundary, the SW material is approximated by an ideal gas, whose density and pressure are given by
\begin{equation}
    \rho_{SW} = \frac{\dot{M}}{4 \pi r^2 V_\infty},
\end{equation}
\begin{equation}
    P_{SW} = k_B T_{SW} \frac{\rho_{SW}}{m_H},
\end{equation}
where the temperature was approximated by $T_{SW} \sim 10^6$~K~\citep{nugis_WRwind_temp,Toalá_2017}, for a simple hydrogen gas approximation.
Given that the flow presents spherical symmetry, {\fontfamily{cmtt}\selectfont PLUTO}'s smoothing function for the initial condition was applied.

\subsection{Nonthermal electrons}
\label{sec:elec}

The strong bow shock formed from the collision of the jet with the SW is considered an efficient particle accelerator~\citep{araudo_2013}. Based on diffusive shock acceleration theory, it is assumed that the electron energy spectrum will follow a power-law~\citep{MATTHEWS2020101543} with an exponential cutoff~\citep{PROTHEROE1999185} at a maximum energy $E_{\rm max}$. The synchrotron and Inverse Compton (IC) cooling constrains the maximum energy, which is dependent on the local magnetic and radiation fields. Before we detail the electron spectrum and its characteristics, it is necessary to introduce a model for the magnetic and radiation fields in and around the knots.

\subsubsection{Magnetic Fields}~\label{sec:mag}

Assuming equipartition between the nonthermal electron and magnetic energy density, \citet{Goodger_2010} estimated a magnetic field $\sim 222~\mu$G for the knot BX2. Studying the fading in the brightness of the knots AX1C and BX2, \citet{Snios_2019} established an upper limit of $80~\mu$G, neglecting acceleration processes. \citet{Sudoh_2020} argues that thermal particles dominate even in the knot region, implying a relatively small magnetic field, $\sim 40~\mu$G, even in the amplified region of the knot. When performing RHD simulations, as is done in this work, it is assumed that any dynamical effect of the magnetic field can be neglected. Comparing the $\sim220~\mu$G magnetic field obtained for BX2 with the amplified pressure on the shock from the simulations ($\sim10^{-8}$ dyn~cm$^{-2}$), gives a ratio between the magnetic to the thermal pressure of $\sim0.1$ at BX2. Since the pressure is higher, the ratio should be lower for the inner knots.

In our modelling, we estimate the strength of the magnetic field on the grid from the thermal pressure as $U_{\rm mag} = P/\beta_m$~\citep{Gómez_1995, Seo_2023}, where $\beta_m$ is considered constant with a value appropriately chosen such that $B$ satisfies the observational constraints in the literature. The value of $B$ can be written as 
\begin{equation} \label{eq:B_P}
    B = \sqrt{\frac{8 \pi P}{\beta_m}} =: \frac{B_P}{\sqrt{\beta_m}},
\end{equation}
where $B_P$ is obtained from the RHD model at the shock position (Table~\ref{tab:knots_estimations}). Combining $B_P$ with the literature's estimations ($\sim 40-80~\mu$G), we found $\beta_m = (B_{P}/B_{\rm knot})^2$, with values $\beta_m^{AX1} \sim 200 - 810$, $\beta_m^{AX2} \sim 160 - 670$, and $\beta_m^{BX2} \sim 50-210$. Assuming $\beta_m$ to be constant along the jet, the maximum magnetic field of the knots $B_{\rm max} = 80~\mu$G gives the constraint $\beta_m > 210$.

\begin{figure}
    \centering
    \includegraphics[width=\linewidth]{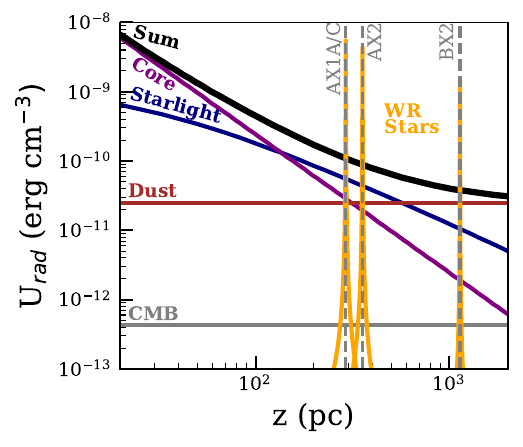}
    \caption{Radiative energy density along Cen~A's jet, used to compute the inverse Compton radiation losses and spectrum. The core, starlight, dust, CMB, and WR stars components are shown (see text for details). The knot locations are shown with vertical dashed grey lines.}
    \label{fig:Urad}
\end{figure}

\begin{table*}
    \centering
    \begin{threeparttable}
    \caption{Radiation and magnetic fields at the positions referent to knots AX1, AX2, and BX2.}
    \begin{tabular}{c|c|c|c|c|c}
        Knot & $U_{0} z^{n}$ \tnote{a}& $U_{\rm rad} = U_0 z^{n} + U_{WR}$\tnote{b} & $B_P~(\mu G)$\tnote{c} & $\beta_m$\tnote{d}\\
        \hline
        AX1 & $1.03\times10^{-7} z^{-1.23}$ & $1.1 \times 10^{-10}$ & 1150 & $200 - 810$\\
        AX2 & $7.42 \times 10^{-8} z^{-1.18}$ & $8.4 \times 10^{-11}$ & 1020 & $160 - 670$\\
        BX2 & $2.86 \times 10^{-9} z^{-0.68}$ & $2.6 \times 10^{-11}$ & 595 & $50 - 210$ \\ 
    \end{tabular}
    \label{tab:knots_estimations}
    \begin{tablenotes}
        \item[a] Local analytical expression for the radiation field around the high $z=z_{\rm knot}$ (Sections~\ref{sec:rad} and \ref{sec:emax}).
        \item[b] Total radiation field at the shock, $z-z_{knot} \approx 0.1 R_j$ from the WR position (Sections~\ref{sec:rad} and \ref{sec:emax}).
        \item[c] $B_{P}$ obtained from the pressure at the shock position (Section~\ref{sec:mag}).
        \item[d] $\beta_m$ based on the literature's estimations for the magnetic field (Section~\ref{sec:mag}).
    \end{tablenotes}
    \end{threeparttable}
\end{table*}

\subsubsection{Radiation Fields}~\label{sec:rad}

Different radiation fields have been identified and estimated along the jet of Cen~A. Figure~\ref{fig:Urad} shows the radiative energy density along the jet expected from the galactic (core, starlight, and gas/dust lane), WR star, and CMB components.

Close to the central AGN, the dominant radiation field comes from the synchrotron and Synchrotron Self Compton (SSC) emitted from the core. In this work, our model for the core radiation follows the SSC fit and spatial dependence from \citet{Tanada_2019}, with a luminosity $L_{\rm nuc} = 1.7 \times 10^{42}$~erg s$^{-1}$.

At distances of hundreds of parsecs from the core, the energy density from the galactic starlight becomes dominant. According to \citet{Tanada_2019}, the spectral distribution of the starlight component is well described by a blackbody with temperature $4 \times 10^3$~K and luminosity $3 \times 10^{44}$~erg~s$^{-1}$.

In addition to the core and starlight radiations, Cen~A presents a third relevant galactic component, which will be denoted dust radiation. Optical images of Cen~A reveal an extended kpc-scale disk of gas and dust, known as the dust lane, possibly associated with a previous galactic merger~\citep{struve_2010}. Based on LABOCA measurements and archival ISO-LWS data, \citet{laboca_2008} extracted the SED from a $1.4$~kpc aperture around the center of Cen~A. We approximate the SED dust as a $30$~K blackbody with luminosity $L_{\rm dust} \approx 10^{44}$~erg~s$^{-1}$. We don't find significant differences in our results using the modified black body distribution $\propto B_{\nu}(T_{1,2}) \nu^2$, applied by \cite{hess2020resolving}, where $B_{\nu}(T)$ is the Planck function, and $T_{1,2} = 14$~K and $30$~K, respectively. Given the extension of the dust lane and since the radius of the LABOCA measurements overcomes the distance of the knots from the core, in this work the dust radiation energy density is approximated as a constant up to the observation radius of the LABOCA measurement, $R_{\rm dust} = 1.4$~kpc, such that the energy density is given by $U_{\rm dust} \sim L_{\rm dust} / 4\pi R_{\rm dust}^2 c$.

The usual cosmic microwave background (CMB) and the radiation field from the WR star might also affect the electrons and their radiation. The WR star is taken as a blackbody of temperature $T_* = 5 \times 10^4$~K and radius $R_* = 2.5 R_{\odot}$ typical for Wolf-Rayet stars~\cite[and references therein]{wykes_2015,araudo_2013}. The energy density in the WR radiation field is approximated by $U_{WR} = \frac{1}{c} \sigma T_*^4 \Big(R_* / D \Big)^2 \approx 3.76 \times 10^{-11} / D_{\rm pc}^2$, where $D_{\rm pc} = D / 1$~pc is the distance from the star's surface measured in parsecs, and $\sigma = 5.67 \times 10^{-5}$~erg~cm$^{-2}$~s$^{-1}$~$K^{-4}$ is the Stefan-Boltzmann constant.

%\hl{The spectrum of the radiation fields considered in the analysis at the different knots can be found in Appendix(Figure}~\ref{fig:Urad_spec})???

\subsubsection{Maximum energy of electrons}~\label{sec:emax}

For known magnetic and radiation fields and shock parameters, $E_{\rm max}$ can be calculated locally at each point along the shock. The maximum energy of electrons is reached when the acceleration time scale is  equal to the energy-loss time scale. We assume diffusive shock acceleration with a coefficient $D = \eta D_B$, where $D_B$ is the Bohm diffusion coefficient describing diffusion with a mean free path equal to the Larmor radius, $r_g$. $\eta$ is a parameter $\geq 1$ that characterises how close the particle acceleration is to the idealised Bohm regime. Our estimate for the acceleration timescale is then given by
\begin{equation}
    \tau_{\rm acc} = \frac{\eta D_B}{v^2 \cos^2 \theta} = \frac{c}{3} \frac{\eta  r_g}{v^2 \cos^2 \theta} = \frac{\eta t_{\rm acc}}{\beta_{\rm ahead}^2 \cos^2 \theta B_{\mu \mathrm{G}}} E_{\mathrm{GeV}} \, ,
\end{equation}
where $D_B$ is the Bohm diffusion coefficient, $\beta_{\rm ahead}=v/c$ is the velocity ahead of the shock in speed of light units, $\theta$ is the angle between the shock normal and the jet velocity, and $t_{\rm acc} = 1.17 \times 10^{-6} \mathrm{\ yr}$. The factor $\cos^2 \theta$ represents an approximate way to account for the combined effects of the velocity projection and the acceleration efficiency likely dropping with shock obliquity~\citep{Caprioli_2014}.

\begin{figure}
    \centering
    \includegraphics[width=0.7\linewidth]{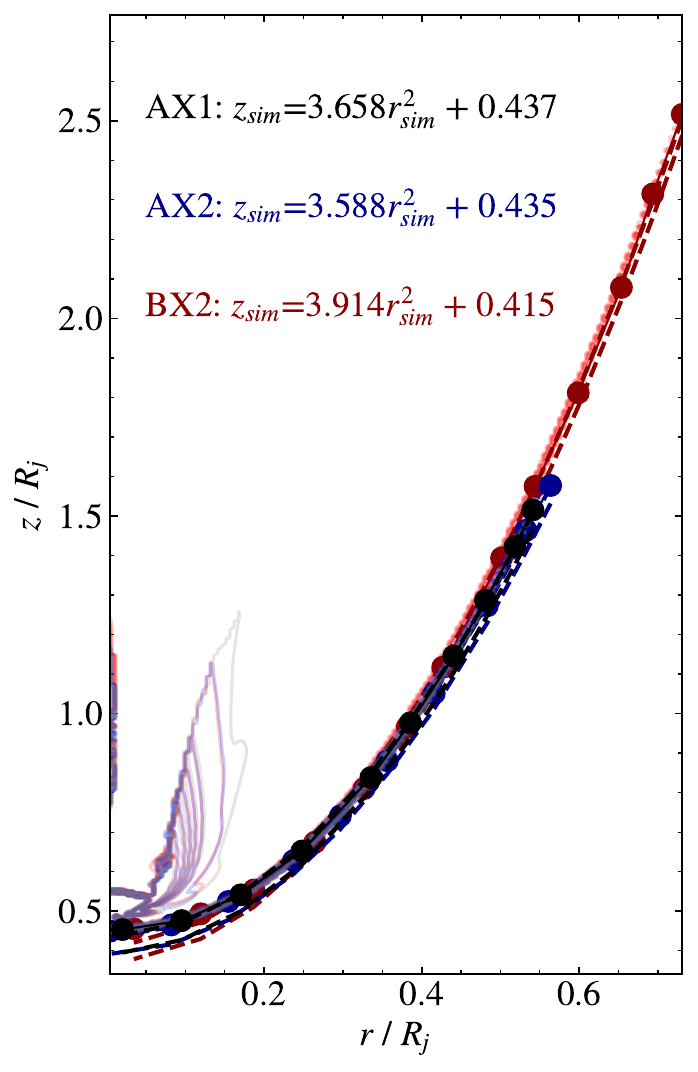}
    \caption{Jet shock for each knot (points) with parabolic fit (dashed line). The parameters obtained in the parabolic fit of each region are shown in the respective colors for each knot: black (AX1), blue (AX2), and red (BX2). The shock structures are the same as shown in the simulation snapshots in Fig.~\ref{fig:RHD:knots}.}
    \label{fig:shocks_fit}
\end{figure}

Given that the shock in the jet material becomes stationary, the shock geometry can be analytically described by a parabola, $z_{\rm shock} = a r^2 + b$, and the angle between the shock normal and the jet velocity field ($\propto \hat{z}$) is given by $\theta = \tan^{-1} (2 a r)$. Figure~\ref{fig:shocks_fit} presents the shock structure and the parameters $a$ and $b$ obtained by fitting the shock profile from the RHD simulations.

The synchrotron and IC cooling timescales are
\begin{align}
    & \tau_{\rm sync} = \frac{t_{\rm sync}}{B^2_{\mu \mathrm{G}}} \frac{1}{E_{\mathrm{GeV}}},\\
    & \tau_{IC} = \frac{t_{IC}}{(U_{\mathrm{rad}}/\mathrm{erg\ cm}^{-3})} \frac{1}{E_{\mathrm{GeV}}},
\end{align}
with $t_{\rm sync} = 1.25 \times 10^{10} \mathrm{\ yr}$ and $t_{IC} = 5.0 \times 10^{-4} \mathrm{\ yr}$. Considering $\tau_{\rm acc} = \tau_{\rm loss} = (\tau_{\rm sync}^{-1} + \tau_{IC}^{-1})^{-1}$, our estimate for the maximum energy reached in the acceleration along the shock is
\begin{equation} \label{eq:emax}
    \frac{E_{\rm max}}{\mathrm{GeV}} = \frac{\chi}{\beta_m^{\frac{1}{4}}} \frac{\beta_{\rm ahead} \sqrt{B_P} }{G\big( z_{\rm shock}(r) \big)^{1/2} } \frac{1}{\sqrt{ \frac{B_P^2}{\beta_m} \big( t_{\rm acc}/t_{\rm sync} \big) + U_{\rm rad} \big( t_{\rm acc}/t_{IC} \big) } },
\end{equation}
with $B_P = \sqrt{8 \pi P}$, $P$ measured in $\mathrm{dyn~cm}^{-2}$ and $G\big( z(r) \big) = ( \cos \arctan \theta(z) )^{-1} = 1 + (2ar)^2$ the obliquity correction. The parameter $\chi \equiv \eta^{-1/2} \leq 1$ accounts for the acceleration efficiency.

For simplicity, in the simulation box around each knot, $U_{\rm rad} (z)$ is approximated by
\begin{align}
    &U_{\rm rad}(r,z) = U_0 z^{n} + U_{WR}(r,z),
\end{align}
with the parameters $U_0$ and $n$ obtained by fitting the total radiation field (without the WR component) for the simulation box of each knot (Table~\ref{tab:knots_estimations}).

Through this work, the cases $\beta_{m} = [210, 670, 800]$, and $\chi = [0.01,1]$, are investigated. Neither of these parameters affects the hydrodynamics, but they are the main parameters that determine the predicted SED from a given simulation.

\subsubsection{Electron population}

After leaving the shock, the electrons injected as a power-law spectrum are subject to synchrotron and IC cooling. Over time, electrons with energy $E$ above the critical energy $E_{b}$ will suffer the effects of cooling, while the population with $E<E_{b}$ remains uncooled, resulting in a broken power law~\citep{longair2011high}. In the stationary state between injection and transport, $E_{b}$ depends only on the position on the simulation grid, such that the electron spectrum in a cell $(i,j)$ is written as
\begin{equation}
    \frac{dN}{dE}\Bigg|_{ij} = A_{ij} e^{-\frac{E}{E_{\rm max,ij}}} \begin{cases}
        \Big(\frac{E}{E_0} \Big)^{-p} \text{,\;\;\;\;\;\;\;\;\;\;\;\;\;\;\;\;\;\;\;\;\;\;\;for } E<E_{b, ij},\\
        \Big(\frac{E_{b, ij}}{E_0} \Big)^{p' - p} \Big(\frac{E}{E_{b, ij}} \Big)^{-p'} \text{, for } E>E_{b,ij},\\
    \end{cases}
\end{equation}
where $A_{ij}$ is proportional to the energy deposited in non-thermal electrons, $U_e$. Assuming $U_e \propto U_B$ and using equation~\ref{eq:B_P}, it is possible to estimate $A_{ij} \propto P_{ij}$. We take $A = A_0 \times P(r,z) \times V(r,z)$, with $V(r,z) = 2 \pi r \delta r \delta z$ the volume of the cell, and $A_0$ a normalization constant. Both $A_0$ and $p$ are determined using the radio/X-ray flux data. Since cooling is synchrotron/IC, $p' = p + 1$.

The critical energy $E_{b}$ results from the temporal cooling of $E_{\rm max}$. The break energy is seeded locally in the shocked cells with initial value $E_b = E_{\rm max}$ and evolves with the fluid in the {\fontfamily{cmtt}\selectfont PLUTO} simulation as a passive scalar or tracer, $Q = \rho^{1/3} / E_{b}$, in a generalization of the method described by \citet{10.1093/mnras/stab3509} for synchrotron cooling in gamma-ray bursts. We adopt a fiducial value of $1$~TeV as the initial maximum energy of the unshocked jet/SW material.

\subsection{Spectral Energy Distribution from radio to gamma rays}
\label{sec:sed}

We evaluated the radio to $\gamma$ ray Spectral Energy Distribution (SED) of the non-thermal radiation emitted by the electron population using the {\fontfamily{cmtt}\selectfont Naima} library~\citep{naima}. Based on equipartition arguments, \citet{Goodger_2010} argues that the origin of the X-ray emission from the knots is mainly synchrotron. \citet{hess2020resolving} also found a synchrotron explanation for the X-rays measured along the jet. Assuming a synchrotron origin for radio and X-ray emission from the knots, the spectral index $p$ of the electron population is given by $p = 2 \alpha_{ph} + 1$. Using the data extracted by \citet{Tanada_2019} for radio/X-ray emission at the knots, the values $p^{AX2} = 2.66$ and $p^{BX2} = 2.27$ were determined. The X-ray emission in knots AX1A/AX1C decreases with increasing energy. In this case, $p$ were estimated using the VLA and the first point of \textit{Chandra} X-ray data, obtaining $p^{AX1A} = 2.40$ and $p^{AX1C} = 2.43$. We assume the same electron population will produce $\gamma$ rays via IC scattering.

We calculate the radio to $\gamma$ ray SED of each knot by summing the nonthermal emission from cells within a radius given by half the \textit{Chandra} resolution ($\sim9$~pc at the distance of Cen~A). In the case of BX2, the knot size found by \citet{Tanada_2019} is larger than the \textit{Chandra} resolution. In this case, the knot is resolved, and so the actual knot size was used to calculate the SED. The curves were normalized to \textit{Chandra} data at $\sim1$~keV, and arbitrarily normalized to the upper and lower uncertainties for $\chi=1$ and $\chi=0.01$, respectively.

\subsection{X-ray images}~\label{sec:xray_image}
To compare the morphology obtained from the model with the X-ray measurements, we built a synthetic X-ray image from our model. The X-ray emission for an observation perpendicular to the jet axis was evaluated as
\begin{equation}
    J_{ij}^{\rm phot} = \sum_{m<i} 2 \frac{j_{mj}^{\rm phot}}{V_{ij}},
\end{equation}
where $J^{\rm phot}_{ij}$ is the total emissivity of the cell after the projection. The emissivity $j^{\rm phot}$ is the emissivity obtained for each cell in the model, normalized using radio/X-ray data, and integrated between $[E_1, E_2] = [0.9, 2.0]$~keV, the range considered in the images of reference~\citep{Snios_2019}. The sum accounts for building a face on view from the simulated emission as expected for cylindrical symmetry. The volume of the cell $V_{ij}$ was considered in correcting the emissivity.

A Gaussian filter was applied to the simulated emission to make for a fair comparison with the \textit{Chandra} images. Comparing to the images from \citet{Snios_2019}, the standard deviation was taken equal to the smooth area used of 3 pixels $\times~0^{\prime\prime}.123$~pixel$^{-1}$. Each pixel of simulation has a size $R_{j} \times \delta r \times 1^{\prime\prime}/18$~pc, implying a standard deviation in simulation units given by $1328.4/ (R_j/{\rm pc})$. The estimation of the knot radius was made according to \cite{Kraft_2002,Tanada_2019} and assuming a standard deviation of $0.5^{\prime\prime}$.

\begin{figure*}
    \centering
    \includegraphics[width=\linewidth]{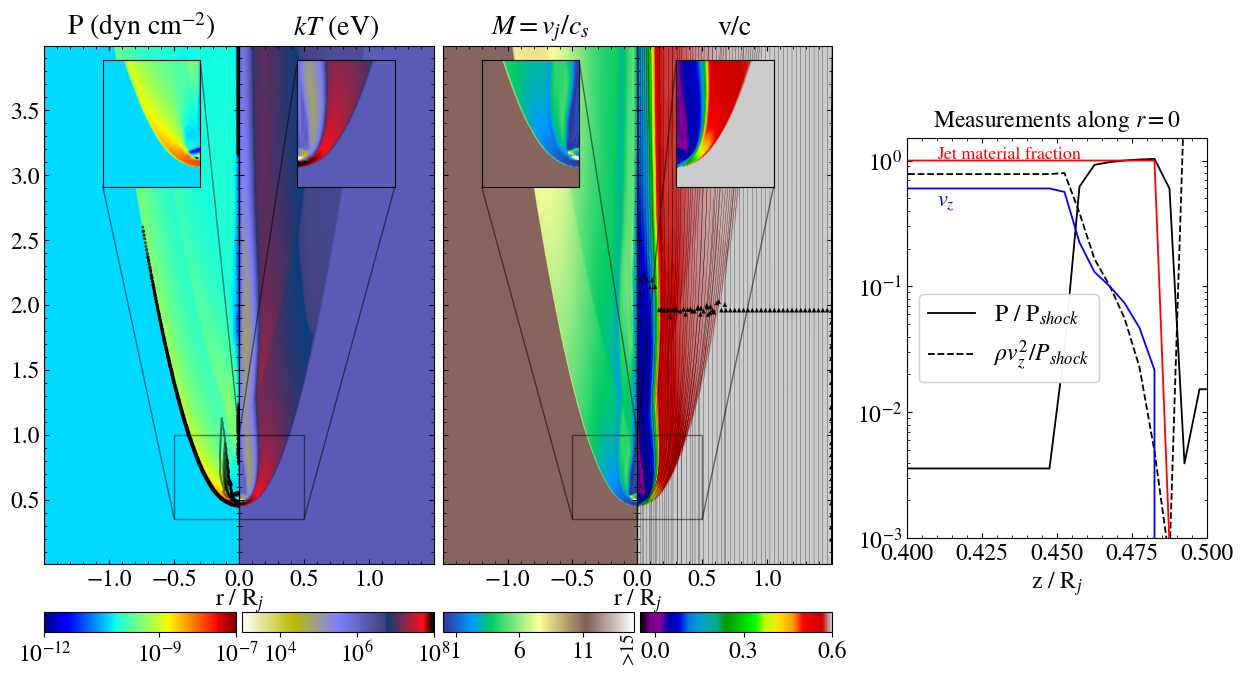}
    \caption{Hydrodynamic properties obtained for knot at position BX2. {\sl Left:} Pressure profile superposed with the shock tracer (black) and $kT$. {\sl Center:} Mach number and velocity profile with velocity streamlines. {\sl Right:} Pressure and ram pressure along the $z$-axis for $r=0$; the fraction of jet material is also shown with a red line.}
    \label{fig:RHD:BX2}
\end{figure*}

\begin{figure*}
    \centering
    \includegraphics[width=\linewidth]{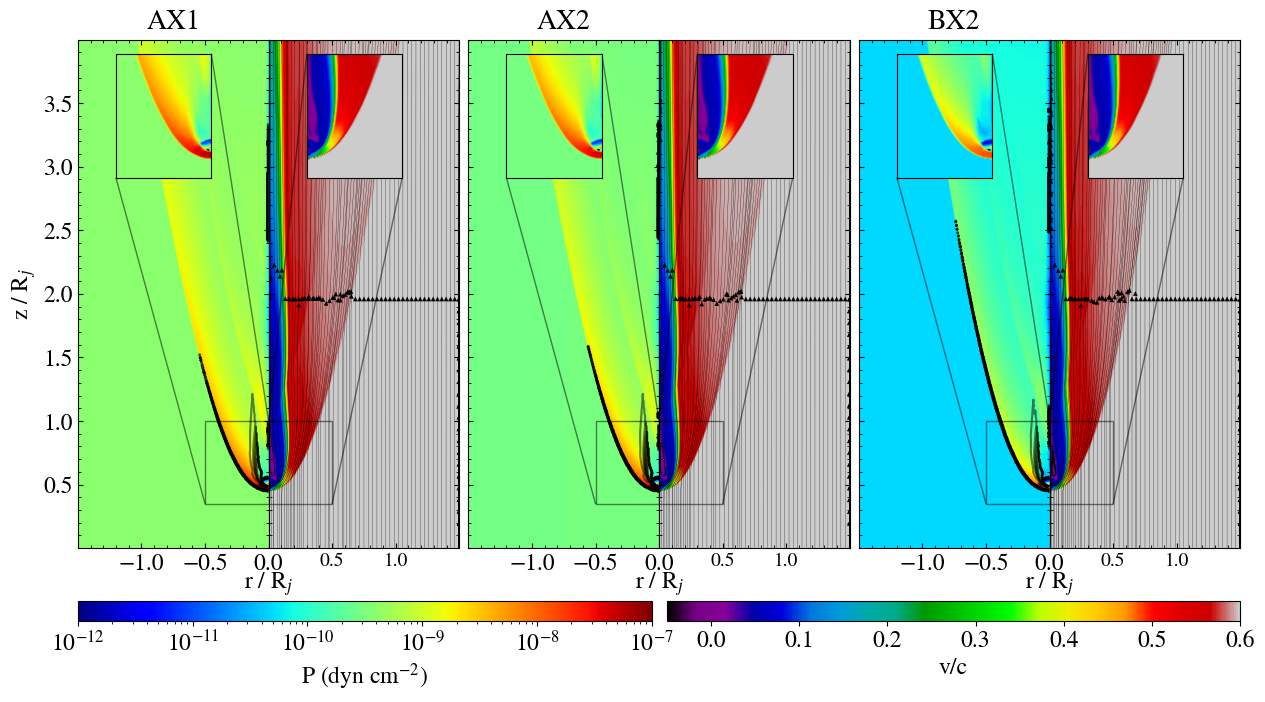}
    \caption{Comparison between knots at position AX1, AX2 and BX2, from left to right. {\sl Left panels:} Pressure profile superposed with the shock tracer (black). {\sl Right panels:} speed and velocity streamlines (black). We observe a similar overall structure to the jet-SW interaction. There is a drop in jet velocity with $z$ and slight changes in bow shock shape (see text and Fig.~\ref{fig:shocks_fit}).}
    \label{fig:RHD:knots}
\end{figure*}

\section{Model results}
\label{sec:model_results}

\subsection{Hydrodynamical structure of knots}
The hydrodynamic quantities obtained for the BX2-like knot are shown in Figure~\ref{fig:RHD:BX2}. The left plot shows the pressure profile, with the double-shock structure superposed, and the temperature $kT$, assuming a hydrogen plasma. A cell is flagged as lying within a shock if $\nabla \cdot \mathbf{v} < 0$ and $\nabla P / P > \epsilon$, where $\epsilon$ is an arbitrarily chosen minimum shock strength threshold. The standard value ($\epsilon = 5$) used in {\fontfamily{cmtt}\selectfont PLUTO} was used and found to well characterise the double shock structure. The central panel shows the Mach number and the velocity profile. The pressure, ram pressure, jet material fraction, and velocity along the $z$-axis are shown in the right panel.

The double shock structure formed around the wind injection region is clearly visible. The jet shock or bow shock is formed at $z_{\rm sim} \approx 0.45$ which can be seen as an intense jump in pressure. At this point, the upstream velocity decreases and, consequently, so does the ram pressure. At $z_{\rm sim} \approx 0.477$ the ram pressure goes to zero and the contact discontinuity is formed. The SW shock starts at $z_{\rm sim} \approx 0.48$. Close to the SW injection region ($z_{\rm sim}=0.5$), the pressure follows a $(z-z_{\rm knot})^{-2}$ profile.

The plasma heats considerably after crossing the jet shock, from $kT \sim1$~MeV to $kT\sim100$~MeV. The heated and highly pressurized plasma flows along the velocity streamlines and is distributed throughout the jet material while it cools. The central region with material cooler than its surroundings corresponds to the SW material, as seen in the velocity field lines. Both the jet and stellar shocks are strong, with Mach numbers $\sim 11$ and $\sim 6$, respectively.

Figure~\ref{fig:RHD:knots} shows the pressure and velocity obtained for knots at positions AX1, AX2, and BX2. The decrease of the jet pressure with $z$ can be seen by comparing to the region upstream of the shock. The jet shock in BX2 is more extended and less radially distributed (see Figure~\ref{fig:shocks_fit}) than the AX-knots, which can be attributed to an increase in the shock intensity with increasing distance from the core. Considering $v_j$ a constant along the jet (which is reasonable in the scale that has been considered in this paper (see \citealt{wykes1d}), and a cylindrical jet), and taking $\Delta P \approx \rho_j \Gamma_j v_j^2 \propto R_j^{-2}(z)$, then $\Delta P / P_j \propto z^{1.5} R_j^{-2}(z)$, implying $(\Delta P / P_j)|_{BX2} \sim 2(\Delta P / P_j)|_{AX1}$.

\subsection{Maximal and break energies}

\begin{figure}
    \centering
    \includegraphics[width=\linewidth]{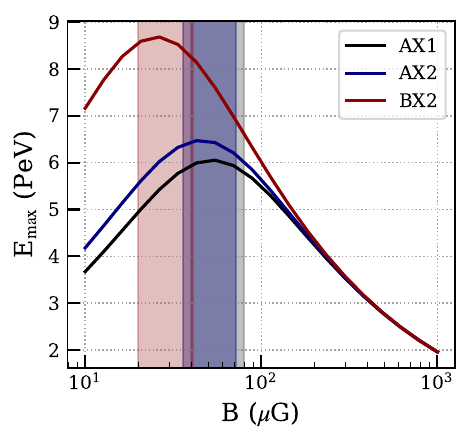}
    \caption{Maximal electron energy for models of knots at positions AX1, AX2, and BX2 as a function of the magnetic field strength, for $\theta=0$ and $U_{\rm rad}$ of Table~\ref{tab:knots_estimations}. For each knot, the colored bands represent the range of magnetic field strengths for $\beta_m \sim 200-800$.}
    \label{fig:emax_B}
\end{figure}

Given the initial plasma conditions and the radiation fields, $E_{\rm max}$ depends on the free parameters $\beta_m$ and $\chi$. Solving equation~\ref{eq:emax}, figure~\ref{fig:emax_B} shows the maximal energy obtained at the position of the knots as a function of the magnetic field intensity for $\chi=1$ and $\theta=0$. In the expected range of values $\beta_m \sim 200-800$, the variations of $E_{\rm max}$ with $\beta_m$ are small, and $E_{\rm max}$ becomes approximately dependent only on $\chi$. The maximum values obtained in the simulations are $E_{\rm max}^{AX1} \sim 3.1$~PeV, $E_{\rm max}^{AX2} \sim 3.4$~PeV, and $E_{\rm max}^{BX2} \sim 4.4$~PeV. The maximum energy increases with the distance from the core, as the energy density of galactic radiation field components decreases downstream the jet. For comparison, the Hillas energy in all knots considered is $0.1-0.3$~EeV, for $\beta_m \sim 200-800$, indicating that the values $E_{\rm max}$ above are achievable and reasonable.

%Figure~\ref{fig:Ebreak_BX2} compares the break energy obtained for different parameters and knots. The strong dependence of $E_{\rm max}$ with $\chi$ becomes very evident (left), as well as the weak dependence on $\beta_m$ (center). The slightly different shock structure of the knots also reflects different distributions of maximum energy along the jet (right).

\begin{figure*}
    \centering
    \includegraphics[width=0.8\linewidth]{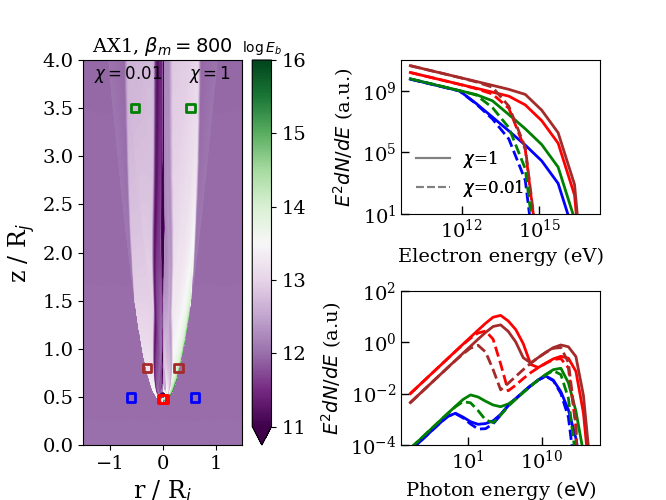}
    \caption{Break energy ($E_b$) and SED in different locations along the AX1 knot, showing the cooling of electrons downstream of the shock acceleration region. The emission is proportional to $p \times V_{\rm cyl}$ of each cell and $A=1$. The left hand panel shows a color map of $\log E_b$ for two different values of $\chi$, with four locations marked with coloured squares. The right-hand panels show, at the location of each coloured-square, the electron spectra ({\sl top right}) and photon spectrum ({\sl bottom right}). Both are plotted in arbitrary units of $E^2 dN/dE$ for the same two values of $\chi$. The SED in the jet region (blue), where $E_{\rm max} = 1$~TeV is shown for completeness, since a fiducial value needs to be put in the definition of $Q=\rho^{1/3} / E_{b}$; however, this region will not be used in the calculations, being showed just for comparison.}
    \label{fig:sites:AX1_by_site}
\end{figure*}

Figure~\ref{fig:sites:AX1_by_site} shows, for the AX1 model, the spatial distribution of the break energy $E_{b}$, together with the electron distribution and the predicted nonthermal radiation at different positions along the simulation box. The cooling of the electrons advected along the velocity streamlines is evident and reflects in the electron spectra. Magnetic field variations result in significant differences in the intensity of the synchrotron peak of radiation compared to the IC peak at different locations. As expected, the value of $\chi$ strongly influences the highest energy radiation in both synchrotron and IC emission. Dust and starlight components are the relevant radiation fields in all cases (see Appendix~\ref{sec:app:radiation_fields}).

\subsection{Nonthermal radiation from the knot centroid}

\begin{figure*}
    \centering
    \includegraphics[width=0.9\linewidth]{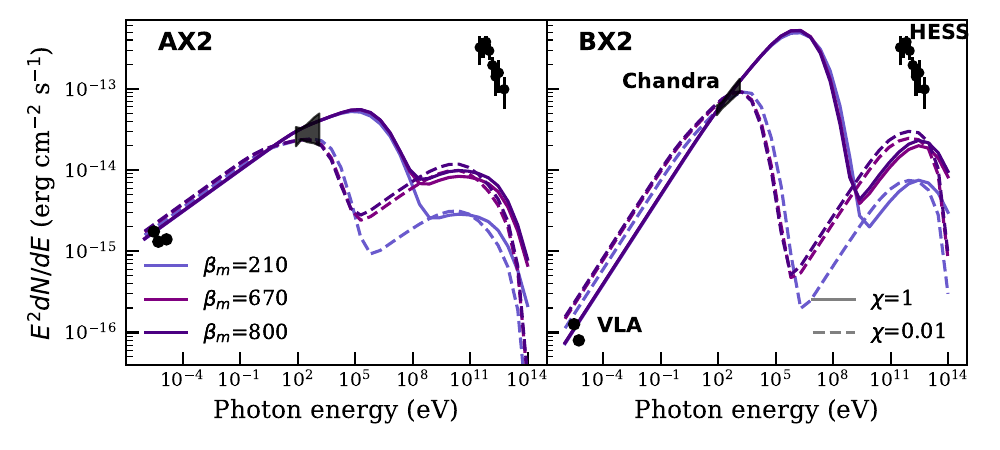}
    \caption{SED predicted using $U_{\rm dust}$ and $U_{\rm starlight}$ for the AX2 ({\sl Left}) and BX2 ({\sl Right}) knots centroid. Circles and butterfly regions show the measurements of VLA, \textit{Chandra}, and HESS. Different colors represent different values for $\beta_m$, and the solid and dashed lines denote $\chi=1$ and $\chi=0.01$, respectively.}.
    \label{fig:sed:AX2_BX2}
\end{figure*}

Figure~\ref{fig:sed:AX2_BX2} shows the SED obtained for the knots AX2, and BX2. The radio and X-ray measurements are satisfactorily reproduced in all scenarios considered. In both cases, the results for $\chi=0.01$ indicate that electrons with energies at least $\sim 0.01$~PeV are necessary to describe the data. This constrains the maximum energy of the electrons to $E_{\rm max} > 10$~TeV ($\gamma \sim 2\times10^7$), assuming a synchrotron origin for the X-ray flux measurements of \textit{Chandra} in agreement with the results obtained by \citet{hess2020resolving}.

\begin{figure*}
    \centering
    \includegraphics[width=0.9\linewidth]{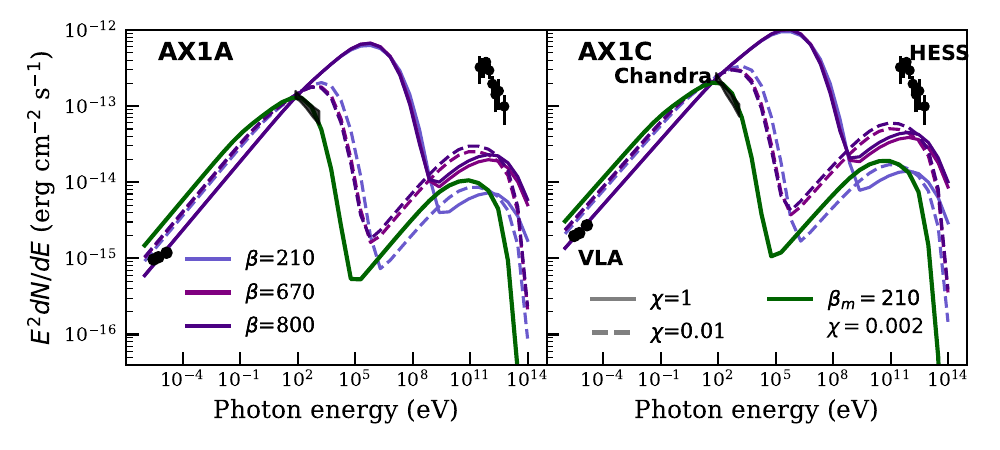}
    \caption{SED predicted using the $U_{\rm dust}$ and $U_{\rm starlight}$ for the AX1A ({\sl Left}) and AX1C ({\sl Right}) knots centroid. The same elements of Figure~\ref{fig:sed:AX2_BX2} are shown, but with an additional SED for $\beta_m=210$, $\chi=2 \times 10^{-3}$ also plotted.}
    \label{fig:sed:AX1}
\end{figure*}

The results for the models of AX1A and AX1C are shown in Figure~\ref{fig:sed:AX1}. Even considering the proposed lower acceleration efficiency, $\chi=0.01$, it is impossible to explain the fading X-ray emission in knots AX1A and AX1C. To achieve an approximate match, we found it was necessary to significantly reduce the maximum energy, requiring $\chi = 2 \times 10^{-3}$; this improves the conformity to the X-ray data, but then overestimates the radio emission from the region by a factor $\sim2$. The result suggests that the acceleration efficiency for the AX1 complex is lower compared to the downstream knots. Possible reasons for this less effective shock acceleration include a more relativistic shock caused by the higher jet velocity, or changes in the jet magnetisation or magnetic field geometry.
%, or a knot origin other than the collision with stars.

\subsection{X-ray morphology}

\begin{figure}
    \centering
    \includegraphics[width=\linewidth]{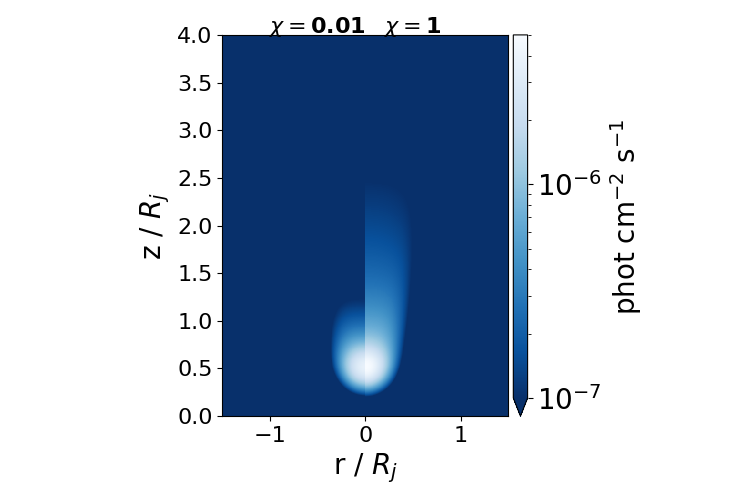}
    \caption{X-ray image for the model of knot BX2 with $\beta_m=210$ for $\chi=0.01$ (\textit{left}) and $\chi=1$ (\textit{right}). For improved clarity, colors are shown on a logarithmic scale, truncated at $10^{-7}$~ photons cm$^{-2}$~s$^{-1}$. The theoretical predictions have been smoothed using a 3 pixel rms Gaussian, as described in section~\ref{sec:xray_image}.}
    \label{fig:BX2_xray}
\end{figure}

Figure \ref{fig:BX2_xray} shows the X-ray morphology obtained for the BX2 knot (similar profiles were obtained for the other knots), for two values of $\chi$. The X-ray signal is stronger at the knot centroid and fades with increasing distance from it. The bright intensity and spatial distribution are dependent on $\chi$, reflecting the SED profile of figure~\ref{fig:sed:AX2_BX2}. A lower maximum energy implies a flux suppression at the end of the considered X-ray range, resulting in a lower integrated number of photons. Also, energy losses remove the electrons from the X-ray emitting range, resulting in a less extended emitting tail.

The qualitative profile obtained from the simulation is compatible with the X-ray images of the knots, as shown in Figure~\ref{fig:cenA} (see also~\citet{Snios_2019}). The predicted number of photons per unit area per unit time is $N_{\rm max}^{BX2} \sim 10^{-6}~{\rm cm}^{-2}~{\rm s}^{-1}$, one order of magnitude higher than reported by \citet{Snios_2019} ($N_{\rm max,obs}^{BX2} \sim 10^{-7}~{\rm cm}^{-2}~{\rm s}^{-1}$), based on a visual inspection. The maximum number of photons observed in the other knots is $N_{\rm max} \sim 10^{-7}~{\rm cm}^{-2}~{\rm s}^{-1}$, keeping a comparable excess of one order of magnitude to the observations.

Using X-ray data from \textit{Chandra} between $\sim0.5 - 8$~keV, \citet{Tanada_2019} estimate the width of each knot perpendicular to the propagation direction. They assume the knot size equals the full width at maximum height (FWMH) corrected by the \textit{Chandra} PSF. The FWMHs obtained in our models, after accounting for \textit{Chandra} resolution, are $\sigma_{AX1}=6.9$~pc, $\sigma_{AX2}=9.2$~pc, independent of $\chi$, and $\sigma_{BX2}=11.8-13$~pc for $\chi=0.01-1$. Comparing these values to the obtained by \citet{Tanada_2019}~(Table \ref{tab:knots_data}), the model provides values larger by factors $\sim 25~\%$, $\sim 5~\%$, and $\sim 4-15~\%$, for AX1C, AX2, and BX2, respectively.
%The FWMHs obtained in our models, after accounting for \textit{Chandra} resolution, are $\sigma_{AX1}=6.5$~pc, $\sigma_{AX2}=7.3$~pc, and $\sigma_{BX2}=8.1$~pc, independent of $\chi$. Comparing these values to the obtained by \citet{Tanada_2019}~(Table \ref{tab:knots_data}), there is a difference of a factor $\sim 20\%$.

The differences found between our model and the \textit{Chandra} measurements can be explained by a slightly weaker stellar wind combined with a larger electron dispersion than that considered in the model. We attribute this to the diffusion of high-energy electrons in the region dominated by stellar material, where the diffusion dominates over the advection. Diffusion will increase the emissivity of these cells, relieving the electron energy density necessary to normalize the X-ray flux (see section~\ref{sec:diffusion}).

The exception is AX1A, where the difference in knot size reaches $\sim 150~\%$. This could indicate that the knot AX1A is formed by a star with a weaker wind. However, VLBI measurements of AX1A~\citep{Tingay_2009} indicate the presence of substructures for the radio counterpart of AX1A, which could require an alternative hypothesis for the origin of  AX1A. One hypothesis is that the interaction between the shocks of the knots in the AX1 region could generate substructures.

\section{Gamma-ray emission from particles accelerated at knots}
\label{sec:extended}

Having so far focused on the hydrodynamics of the jet-SW interactions and the radio and X-ray properties, we now turn our attention to the role of knots in producing VHE $\gamma$-ray emission. \citet{Sudoh_2020} raised the hypothesis that the electrons accelerated at the knots could energize the entire jet, which is supported by our result for $E_{\rm b}$ as shown in Figure~\ref{fig:sites:AX1_by_site}. The energy of the electrons at the top of the simulation box is $\sim 10^{13}$~eV, and the losses must decrease far from the galaxy. The emission of nonthermal radiation is not restricted to the knot centroid, but the entire perturbed/post-shock region contributes to high-energy radiation, in particular $\gamma$ rays. Electrons are accelerated in shock waves up to at least $10$~TeV and cool as they travel along the jet. The synchrotron component could explain the diffuse X-ray emission found in the jet, and the IC could contribute to the signal detected by the Fermi-LAT satellite and HESS Observatories.

Limited angular resolution prevents current $\gamma$-ray experiments from resolving the emission from individual knots. Thus, in addition to the AX2 and BX2 knots centroid and extended emission, other X-ray knots may contribute to the $\gamma$-ray flux measured by the Fermi and HESS observatories. Up to 2010, 40 X-ray knots have been detected by \textit{Chandra}~\citep{Goodger_2010}, and in the 2024 analysis, others have been resolved~\citep{Bogensberger_2024}.

Knots are classified according to their distance from the centroid and labeled from A to G along the jet, and S in the counterjet~\citep{Goodger_2010} (see also Figure~\ref{fig:cenA}). The X-ray knots are mainly distributed in a region between $\sim 15 - 30'' \sim 270-540$~pc away from the nucleus, which we will refer to as the AX-complex. This complex contains the X-ray knots AX1A, AX1C, AX2, AX2A, AX3, AX4, AX4A, AX5, and AX6. The BX-complex is formed by the X-ray knots BX1, BX2, BX2A, BX3, BX4, and BX5. The knots AX1A, AX1C, and BX2 are the brighter knots at 1 keV, followed by AX6 and AX2 \citep{Goodger_2010}. 

Complexes C and E contribute to $\sim 8\%$ and $\sim 1\%$, respectively, of the total X-ray flux that comes from the knots A and B at $1$~keV. Complexes F and G are further away from the centroid. In particular, the G complex, the outermost complex at $\sim 200''$ ($\sim 3.6$~kpc), is located at the infrared flare point, where there is also a local X-ray peak. Downstream, there is an absence of X-ray emission, and so this location has been proposed to be the last point of high-energy particle acceleration along the jet~\citep{hardcastle_infrared_2006}, although we note that acceleration of very-high energy (or even ultrahigh energy) particles in the giant lobes has also been regularly discussed \citep[e.g.][]{hardcastle_high-energy_2009}. 

Due to the large number of X-ray knots, our analysis will focus only on the contribution of the brighter inner complexes A and B. The hydrodynamic models for AX2 and BX2 will be used as a proxy for the knots of complex A and B, respectively. The approximation is expected to be valid, as dust is the dominant radiation field to $\gamma$-ray production in both A and B complexes and pressure variations have a second-order effect (Figure~\ref{fig:RHD:knots}). This approximation ignores interactions between different knots: the perturbed plasma from one knot could collide with another knot, increasing turbulence, for example. 

We selected the AX- and BX- knot complexes since (a) these are the regions containing the brighter knots in X-ray, and we expect this to be reflected in the $\gamma$-ray luminosity; (b) the galactic radiation fields change with the distance and beyond $1.4$~kpc there are no measurements of the dominant dust radiation field; (c) we did not simulate jet collisions far from the BX-complex, which could generate major prediction inaccuracies due to different hydrodynamical conditions. 

To make predictions of the $\gamma$-ray spectrum, we use the available X-ray and radio data to constrain the normalisation and slope of the electron spectrum in each knot. In particular, we use the radio to X-ray spectral index $\alpha_{4.8}^X$ obtained by \citet{Goodger_2010} for each knot to characterize its emission. The electron spectral index is taken as $p = 2 \alpha_{4.8}^X + 1$. The SED was integrated in the same spatial area for the respective proxy, AX2 or BX2. To normalise our spectra, we account for the differences observed in the X-ray flux. After normalizing the knots so that, within observational uncertainties, they produce the same emission at 1~keV, the SED is weighted using the flux density $j_{X}$, at $1$~keV, by $f_i = j_{X}^i / j_{X}^{AX2}$ for AX2-like knots and $f_i = j_{X}^i / j_{X}^{BX2}$ for BX2-like knots. Our overall approach has the advantage that we use observing frequencies (X-ray and radio) at which the knots can be individually resolved to inform our modelling, which is then used to make predictions, integrated across multiple knots, for the much lower resolution $\gamma$-ray waveband.  

The model we have developed for AX2 and BX2 is suitable for knots whose origin can be attributed to jet-SW interactions, with an X-ray spectrum dominated by uncooled electrons ($\alpha_X<1$). For some of the X-ray knots this origin has been discarded. Knots AX3, AX4, and BX2A (and also SX2A and CX4) move at relativistic speeds, ruling out an origin due to a collision of the jet with obstacles~\citep{Bogensberger_2024}. Comparing the proper motions measured in X-ray and radio for knots AX3 and AX4, previously identified as the radio knots A3A and A3B, \citet{Bogensberger_2024} argues that they are not radio/X-ray counterparts. AX2A was detected only in two \textit{Chandra} observations, being unlikely a jet knot~\citep{Bogensberger_2024}, and will not be included in the analysis. Similarly, knots BX3 and BX5 were not present in the 2024 analysis and will also be excluded.

Table~\ref{tab:knots_complex} summarizes the main properties of the X-ray knots used in the analysis. Throughout this section, we will focus on $\beta_m=210$, since in this case $B_{\rm shock}^{AX1} \approx 80~\mu$G, within the maximum allowed value estimated by \citet{Snios_2019}, and $B_{\rm shock}^{BX2} \approx 40~\mu$G, the expected value of \cite{Sudoh_2020}.

\begin{table*}
    \centering
    \begin{threeparttable}
    \caption{Measurements of the X-ray fluxes and spectral indices for the knots discussed in this work.}
    \begin{tabular}{c|c|c|c||c|c}
        Knot & X-ray flux 1 keV (nJy)\tnote{a} & $\alpha_X$ \tnote{a} & $\alpha_X^{4.8}$ \tnote{a} & $p$ & $p$ reference\\
        \hline
        AX1A & $10.65$ & 1.08 & 0.85 & 2.40 & Fitting data from \citet{Tanada_2019}\\
        AX1C & $21.43$ & 1.06 & 0.85 & 2.43 & Fitting data from \citet{Tanada_2019}\\
        \hline
        AX2 & $9.29$ & 0.77 & 0.86 & 2.66 & Fitting data from \citet{Tanada_2019}\\
        AX2A & $3.19$ & 0.56 & - & - & Not a jet knot~\citep{Bogensberger_2024}\\
        AX3 & $4.02$ & 0.78 & 0.80 & 2.6 & Relativistic speed~\citep{Bogensberger_2024}\\
        AX4 & $4.98$ & 0.94 & 0.88 & 2.8 & Relativistic speed~\citep{Bogensberger_2024}\\
        AX4A & $0.30$ & 0.48 & - & - & Taken as AX6\\
        AX5 & $6.60$ & 0.59 & 0.61 & 2.2 & $p = 2 \alpha_X^{4.8} + 1$\\
        AX6 & $9.39$ & 0.51 & 0.70 & 2.4 & $p = 2 \alpha_X^{4.8} + 1$\\
        \hline
        BX1 & $3.69$ & 0.83 & - & - & Taken as BX4\\
        BX2 & $19.39$ & 0.63 & 0.67 & 2.27 & Fitting data from \citet{Tanada_2019}\\
        BX2A & $1.44$ & 0.58 & - & - & Relativistic speed~\citep{Bogensberger_2024}\\
        BX3 & $2.02$ & 1.40 & - & - & High $\alpha_X$; not in \citep{Bogensberger_2024}\\
        BX4 & $4.94$ & 0.91 & 0.77 & 2.5 & $p = 2 \alpha_X^{4.8} + 1$\\
        BX5 & $3.06$ & 1.13 & - & - & High $\alpha_X$; not in \citep{Bogensberger_2024}\\
    \end{tabular}
    \label{tab:knots_complex}
    \begin{tablenotes}
        \item[a] From \citet{Goodger_2010}.
    \end{tablenotes}
    \end{threeparttable}
\end{table*}

\subsection{Knot complexes: centroid emission}
\label{sec:knots_comp_core}

\begin{figure*}
    \centering
    \includegraphics[width=0.9\linewidth]{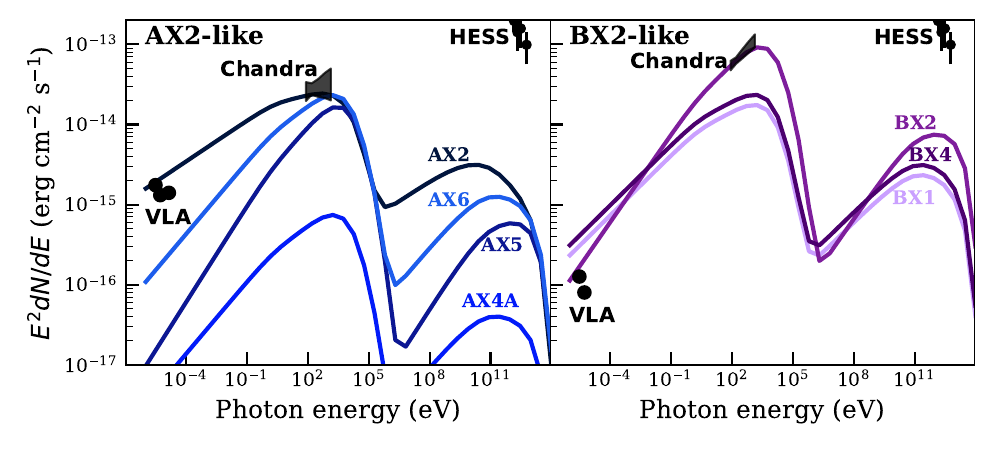}
    \caption{SED predicted for different knot groups for $\chi=0.01$, $\beta_m = 210$. VLA and \textit{Chandra} data are presented for comparison and are taken from \citet{Tanada_2019} for knots AX2 ({\sl Left}) and BX2 ({\sl Right}).}
    \label{fig:sed:all_core}
\end{figure*}

Figure~\ref{fig:sed:all_core} shows the predicted SED for the different groups of knots. The effect of the spectral indices and flux weight can be seen clearly. The predicted $\gamma$-ray flux density between different knots approximately tracks the X-ray flux density, since the same electron population generates both components through synchrotron and IC emission, respectively. The centroid emission of BX2 is dominant compared to that of the other knots. Although the knot centroids produce some $\gamma$-ray, this emission is not sufficient to account for the HESS data.

The radio emission of BX1 exceeds that of BX2 and is of the same order as that for BX4. However, the radio counterpart of BX1 was not detected~\citep{Goodger_2010}, while it was observed both for BX2 and BX4. This fact challenges the applicability of the model for BX1.

\subsection{Knot complex: extended emission}
We now predict the $\gamma$-ray emission from the entire simulation box for all knots, not only the centroid region of each knot. Only the region downstream of the parabolic jet shock was considered, since the model gives no information about the electron distribution (spectral index and maximum energy) in the upstream jet flow, and we expect the emission in that region to be subdominant. The flux is normalized according to the emission in the knot centroid, as discussed in section~\ref{sec:knots_comp_core}.

\begin{figure*}
    \centering
    \includegraphics[width=\linewidth]{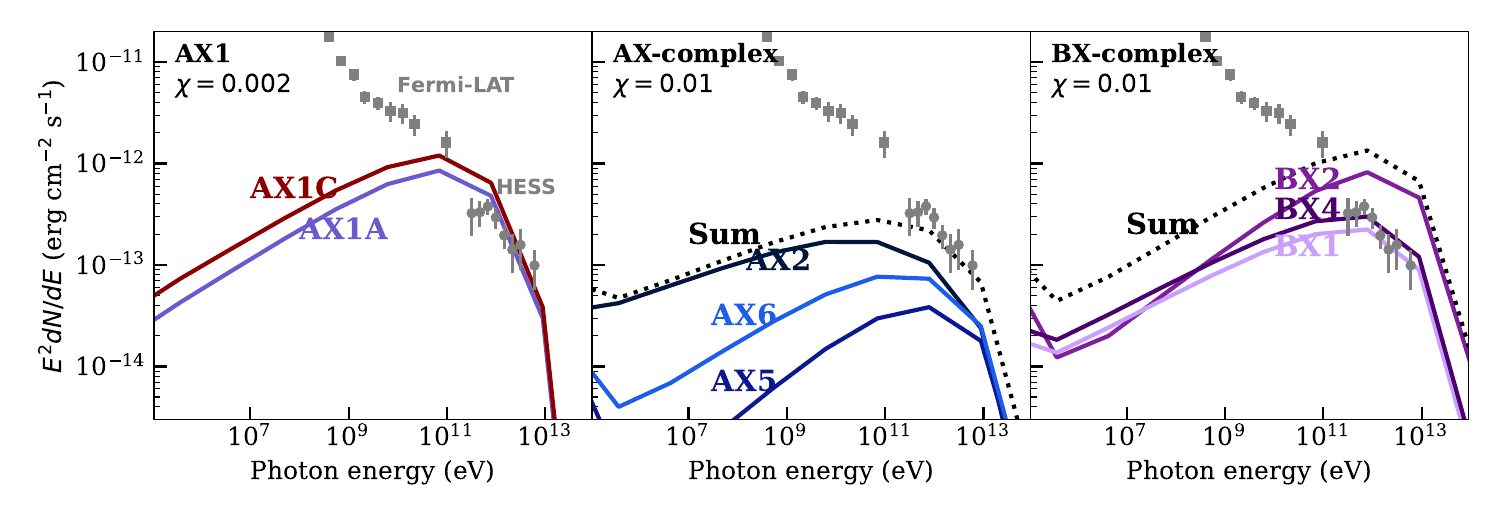}
    \caption{Sum of the knot centroid and extended nonthermal emission predicted for different knot groups for $\beta_m = 210$, with $\chi=0.01$ for AX- ({\sl Center}) and BX-complex ({\sl Right}), and $\chi=2\times10^{-3}$ for AX1A/AX1C ({\sl Left}).}
    \label{fig:sed:all_diffuse}
\end{figure*}

Figure~\ref{fig:sed:all_diffuse} shows the SED obtained for the sum of emission from the knot's centroid and extended region. AX1A, AX1C, and BX2 are the brighter $\gamma$-ray sources in the jet, with BX2 contributing with fluxes at higher energies than AX1A/AX1C. AX2 is the dominant source of $\gamma$ rays in the AX-complex, and the total AX-complex emission agrees with the HESS measurements for $\beta=210$, $\chi=0.01$. In the BX-complex, the $\gamma$-ray flux from BX2 is dominant, and, as in X-ray, the $\gamma$-ray emission from BX2 is the higher between all the knots considered. In general, the emission from the BX-complex overcomes that from the AX-complex, and the emission from BX2 exceeds the $\gamma$-ray flux measured by HESS.

The total energy in the non-thermal emission cannot exceed the available energy provided by the jet and stellar wind bow shocks. For consistency, we compare the non-thermal radiation luminosity of the extended and centroid emission from BX2 ($L_{\rm rad}^{\rm BX2}$) and the available luminosity from the jet bow shock ($L_{\rm jbs}$). According to \citet{araudo_2013}, $L_{\rm jbs} = \left( \frac{R_s}{R_j} \right)^2 L_j \approx 5\times10^{40}~{\rm erg}~{\rm s}^{-1}$, for $R_s \sim 0.05~R_j$ (see Figure~\ref{fig:RHD:BX2}) and $L_j = 2 \times10^{43}~{\rm erg}~{\rm s}^{-1}$. Integrating the non-thermal spectrum of Figure~\ref{fig:sed:all_diffuse} from radio to $\gamma$ rays, we obtain $L_{\rm rad}^{\rm BX2} \approx 1.4 \times 10^{40}$, or a fraction $\sim25~\%$ of the energy available on the shock. Similar calculations for AX1A and AX1C gives $\sim30~\%$ and $\sim50~\%$, respectively. Considering the sum of knots, the energy requirements can be distributed across different shocks, relieving the energetic requirement. For AX-complex, for example, the fraction is $\sim15~\%$ of one jet bow shock, but multiple jet bow shocks are contributing. Although high, these values indicate that the energy dissipated at the shock is enough to account for the non-thermal emission obtained here.

\begin{figure}
    \centering
    \includegraphics[width=\linewidth]{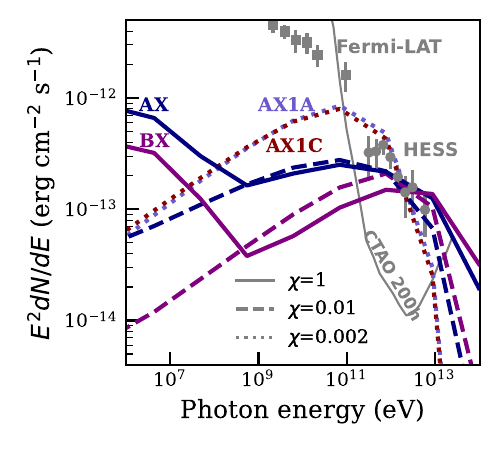}
    \caption{Sum of the knot centroid and extended nonthermal emission predicted for the knots AX1A, AX1C, AX-complex (AX), and BX-complex (BX), and $\beta_m = 210$. The models AX1C and BX-complex are normalized to the HESS data.}
    \label{fig:sed:sum_diffuse}
\end{figure}

The sum of the centroid and extended $\gamma$-ray emission from the combination of knots exceeds the signal measured by Fermi-LAT and HESS independent of the value of $\chi$ even for the high magnetic field considered $\beta_m=210$.
Figure~\ref{fig:sed:sum_diffuse} shows the SED for the knots, for $\beta_m=210$, normalized to the HESS data. Considering the BX-complex a factor $\sim 6.5$, is needed as a normalization factor ($\sim 4$ for BX2 only). AX1A or AX1C only demands a factor $\sim 2$, while if all knots are taken into account, the $\gamma$-ray excess is $\sim 10$. The excess can be explained by an overestimation of the normalization in the knot centroid, due to diffusion, for example, a higher magnetic field than considered (section~\ref{sec:impl:mag}), or a different fraction of non-thermal energy in the knot centroid and extended region. The last hypothesis has already been raised by \citet{Sudoh_2020}.

\subsection{Cooled electrons emission}

The $\gamma$-ray emission from the extended region of the knots can explain the VHE $\gamma$ rays measured by HESS, but is unlikely to describe the excess detected by the Fermi-LAT concerning the core SED. The Fermi peak requires a break energy of the electron spectrum lower than that provided in the knot region. This raises the question whether after propagating through the jet, the electrons are sufficiently cool to explain the Fermi-LAT signal.

\begin{figure}
    \centering
    \includegraphics[width=\linewidth]{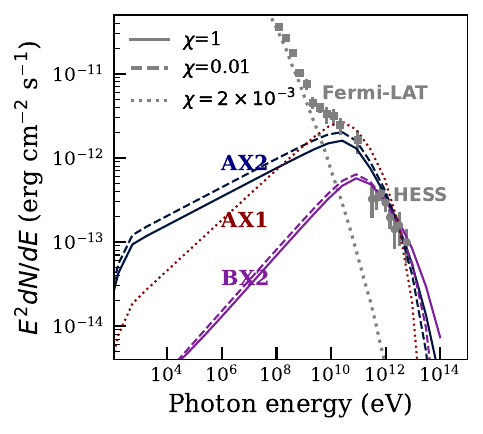}
    \caption{Comparison between the profiles for the $\gamma$-ray measurement by Fermi-LAT and HESS and the prediction for electrons accelerated at the knots AX1, AX2, and BX2 after propagating along the $\sim3.6$~kpc jet extension. The emission model is described in Section~\ref{sec:model_results}, where $\chi$ measures the acceleration efficiency. The curves are arbitrarily normalized to the data points.}
    \label{fig:sed:analytic}
\end{figure}

Considering the initial maximum energy at the knots $\sim 4 \chi $~PeV and assuming the electrons are advected by the fluid at $\beta_j \approx 0.6$, it is possible to estimate the break energy of the spectrum after propagation through the entire jet length, reaching $\sim 3.6$~kpc. There is no information about the radiation fields in the whole jet, so we assumed the dust and CMB components are constant up to $3.6$~kpc (the other radiation fields are expected to be subdominant, see Figure~\ref{fig:Urad}). We focus on particles accelerated at AX1 ($z_{\rm initial} \approx 293$~pc), AX2 ($z_{\rm initial} \approx 356$~pc) and BX2 ($z_{\rm initial} \approx 1134$~pc). The electrons reach the end of the jet with $E_{b} \sim 1$~TeV. The emission of the cooled electrons is normalized to the HESS data, in a qualitative approach to the problem. Figure~\ref{fig:sed:analytic} shows the profile obtained in both cases. Electrons accelerated in the knots provide a good explanation for the HESS emission, but only the cooler electrons, from inner AX1 and AX2 knots explain the data from Fermi-LAT and HESS simultaneously. The possibility is that these electrons are trapped in the lobes where they emit~\citep{fermi_inner_cenA}.

\section{Implications}
\label{sec:implications}

We have shown that particle acceleration at jet knots provides a reasonable explanation for the VHE $\gamma$-rays observed from Cen A on kpc-scales. We have already discussed some of the ways in which our results agree or disagree with the observed data. We now explore further implications of this work for the jet properties, diffusion physics and acceleration of protons and ions to even higher energies.

\subsection{Jet velocity} 
After passing through the bow shock, the jet material is decelerated, as seen in Figure~\ref{fig:RHD:knots}. If the jet widening, observed at $\sim250$~pc from Cen~A's core~\citep{Goodger_2010}, is caused by the interaction with the obstacles that generate AX1A and AX1C, this raises the interesting possibility that the true jet speed could be higher in the inner region than reported. \citet{Snios_2019} compare the apparent jet speeds along the jet for both M87 and Cen~A. The M87 jet presents superluminal motion in the range $\sim 10-1000$~pc, while there is no data for Cen~A in the $\sim 10-100$~pc scale. If the jet of Cen~A accelerates after $\sim 10$~pc as reported to M87, the collision that results in the AX1A/AX1C may make the jet wider and slower. Further support for this scenario comes from \citet{Bogensberger_2024}, who find the AX4 knot is moving at $\sim0.9c$ indicating the existence of significantly higher speeds within the Cen~A jet.

\subsection{Diffusion effects}
\label{sec:diffusion}

The method we used to solve for the electron distribution is suitable when the transport of accelerated particles is dominated by advection, since it does not account for diffusive effects. The timescales for diffusion and advection are
\begin{align}
    & \tau_{\rm dif} = \frac{L^2}{2 D_B} = t_{\rm dif} \frac{L_{\rm pc}^2 B_{\mu \mathrm{G}}}{E_{\mathrm{PeV}}},\\
    & \tau_{\rm adv} = \frac{L}{v} = t_{\rm adv} \frac{L_{\rm pc}}{\beta},
\end{align}
where $L_{\rm pc}$ is a characteristic scale measured in parsecs, $v = \beta c$ is the fluid speed of the region, $t_{\rm dif}=4.88 \mathrm{\ yr}$, and $t_{\rm adv} = 3.26 \mathrm{\ yr}$.

The transition length scale between the dominance of diffusion and advection can be estimated by equating the relevant timescales, which gives
~
\begin{align}
    L_{\rm pc}^{\rm dif/adv} = \frac{2}{3}\frac{E_{\rm PeV}}{\beta B_{\mu G}}.
\end{align}

To estimate the diffusion radius around the shock, we assume $B_{\rm knot} \sim 40 ~\mu$G. The jump condition for the magnetic field obtained in the simulation is $\sim 10-15$. In the jet material around the shock, $B \sim 4~\mu$G, $\beta_j = 0.6$, and $E_{\rm max} \sim 4$~PeV, the maximum displacement due to diffusion is $L_{\rm jet} \sim 1$~pc. There is a small correction to the knot radius due to diffusion. On the other hand, in the SW material, assuming $B \sim 1~\mu$G, $\beta_{WR} = 10^{-2}$, and $E_{\rm max} \sim 4$~PeV, the maximum displacement is $L_{WR} \sim 270$~pc. Compared to the knots' radius, the accelerated electrons could fill the entire stellar material region inside the shock. This behaviour will lead to an increase in the emission area, relieving the problems with the high emissivity obtained in the model and motivating future calculations that account for a more detailed treatment of particle transport. 

Approximating the inner knot region as the shell of a hemisphere, the emitting volume will be $\sim \cfrac{1}{2} \cfrac{4 \pi}{3}( (R + \delta R)^3 - R^3 )$, with $R \sim 0.1 R_j$ the shock radius and $\delta R \sim 0.025 R_j$ the shock width. Estimating that the diffusion will increase the emitting region to the full sphere, of radius one parsec above, the normalization to the radio/X-ray flux in the centroid will drop by a factor $\sim5$. Given the spatial scale of diffusion, this effect can account for the strong peaked emission obtained in the X-ray images of the model, and can increase the radius of the knots.

%However, the electrons will suffer cooling while spread through the jet material. Diffusion will happen without cooling for a distance of approximately $\tau_{cool} = \tau_{\rm dif}$, implying a distance
%\begin{equation}
%    L_{\rm pc}^{dif/cool} = \sqrt{ \frac{B_{\mu G}}{B_{\mu G}^2 t_{\rm dif}/t_{\rm sync} + U_{\rm rad} t_{\rm dif}/t_{IC}} }.
%\end{equation}
%
%Using the $U_{rad}$ evaluated at the shock (Table \ref{tab:knots_estimations}), the maximum displacement without losses during the diffusive motion is $L_{jet}^{dif/col} \sim 2$~pc for all knots.

In these estimates we have considered the idealized Bohm diffusion coefficient as the one relevant for the diffusion process; this is equivalent to setting the diffusion mean free path equal to the Larmor radius. A more realistic scenario should assume $D(E) \approx D_B(E) \Big[ (E/E_d)^{-2/3} + \nicefrac{2}{3} (E/E_d) \Big]$, where $E_d \approx 0.9 B_{\mu G} \lambda_{c, pc} / 2\pi$~\citep{globus_2008}. The coherence length of the magnetic field, $\lambda_{c}$, is hard to estimate. Assuming $\lambda_c \sim (0.1-1)R_j$ will increase the diffusion coefficient, relative to Bohm, by a factor of $2-4$ at the energy of $4$~PeV for $B\sim4~\mu$G and BX2.\footnote{Increasing the diffusion coefficient in the acceleration site will lengthen the acceleration timescale, leading to a less efficient acceleration. This way, writing $D(E) = D_B (E) \eta(E,E_d)$, implies that $\chi = \eta(E_{\rm max},E_d)^{-1/2}$ (see section~\ref{sec:elec}).}

\subsection{Magnetic Fields} 
\label{sec:impl:mag}

We use the VLA/\textit{Chandra} data to define the amplitude of the synchrotron emission in the model. The amplitude of the synchrotron spectrum increases as $B^{(p+1)/2}$, whereas the amplitude of the IC component depends on $B^{-(p+1)/2}$, meaning that the higher the magnetic field, the lower the luminosity of $\gamma$-rays (considering the X-ray amplitude fixed by observations). The comparison of the model to the HESS flux provides a lower limit for the magnetic field in the knots. The high flux obtained for the knots BX2 may suggest a higher magnetic field in the BX-complex than assumed here, or $\beta_m < 210$. To account for the factor $\sim 4$ $\gamma$-ray excess from BX2 (see Figure~\ref{fig:sed:all_diffuse}) a magnetic field of $\sim 90~\mu$G is necessary, closer to the upper limit calculated by \citet{Snios_2019}.

\begin{figure}
    \centering
    \includegraphics[width=\linewidth]{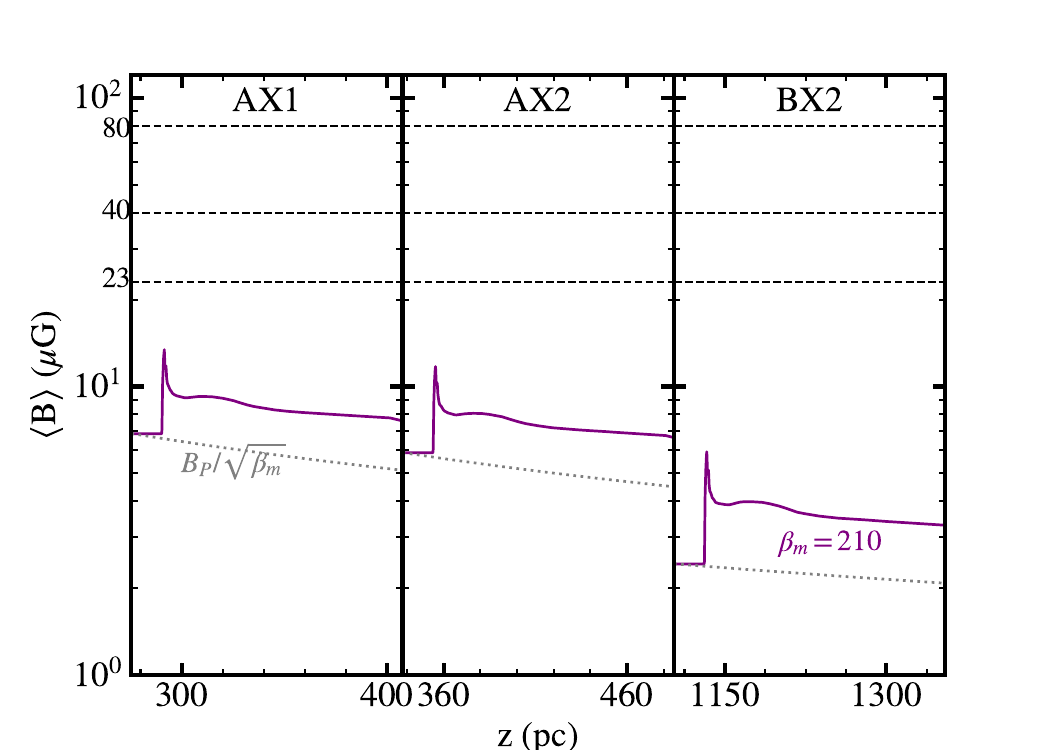}
    \caption{Mean magnetic field along the jet for $\beta_m=210$. The mean values obtained in the simulation box for the knots AX1 ({\sl Left}), AX2 ({\sl Center}), and BX2 ({\sl Right}). The magnetic field intensity obtained from the pressure profile (eq.~\ref{eq:B_P}) is shown in gray. Values obtained in the literature ($40~\mu$G~\citep{Sudoh_2020}, $<80~\mu$G~\citep{Snios_2019}, and $23~\mu$G~\citep{hess2020resolving}) are also shown. }
    \label{fig:B_z}
\end{figure}

Figure~\ref{fig:B_z} shows the radially averaged magnetic field in the simulated regions along $z$ for the different values of $\beta_m$ used in this work. The magnetic field of $23~\mu$G estimated by HESS and the approximate analytic expectation from equation~\ref{eq:B_P} are also shown. The $\sim 23~\mu$G HESS estimate was obtained using the SED data on scales $2.4-3.6$~kpc in projection from \citet{hardcastle_infrared_2006}, which contains knots from the F- and G-complexes. Given that the jet magnetic field should decrease with the distance from the core~\citep{osullivan_2009}, extrapolating this value back along the jet will lead to magnetic fields with unreasonably high values in the inner region, including in the knots. A more probable scenario is that the magnetic field in that region is amplified by the knots, consisting of an effective, perturbed field. This amplification could take place through cosmic ray-driven instabilities at the shock \citep[e.g][]{bell2004}, with subsequent turbulent magnetic field amplification in the disrupted flow behind the jet-SW interaction. 
Considering the values for the magnetic field obtained here and given that the jump in the magnetic field in the shock is $\sim 15$, values $\sim 1~\mu$G for the non-amplified magnetic field in the jet on the $\sim 3$~kpc scale are more likely.

\subsection{Knots as UHECR accelerators ?}
Shock acceleration is among the prime candidate mechanisms to accelerate particles to ultra-high energies~\citep{MATTHEWS2020101543,rieger2022}. In addition to electrons, the shock generated from the collision between the jet and the stellar wind can accelerate protons and heavier nuclei, which cool much more slowly and can have higher maximum energies. \citet{wykes2018uhecr,muller} propose that the wind from a WR star could provide heavier elements, not usually expected to constitute the jet and necessary to explain the UHECR composition inferred from the Pierre Auger Observatory data~\citep{Mayotte:2023Nc}. Posteriorly, these nuclei could be (re)accelerated by jet shock waves, turbulences, the shear layer between the jet and the environment, or in the lobes~\citep[e.g.][]{hardcastle_high-energy_2009,matthews_ultrahigh_2019,rieger2022}.

Assuming acceleration occurs at the knot centroid with size $\sim0.1 R_j$, the maximum energy reached by a nucleus of charge $Z$ will be 
$E_{\rm Hillas} \sim Z (B/ \mu\text{G}) (R_j /\text{pc}) \times 10^{-4}$~EeV \citep{hillas_origin_1984}. 
Taking the upper limit magnetic field estimated by~\citet{Snios_2019}, the maximum energies reached in the knots AX2 and BX2 are $E_{\rm max}^{AX2} \sim 0.2 Z$~EeV and $E_{\rm max}^{BX2} \sim 0.4 Z$~EeV, respectively. This represents an optimistic scenario in which the diffusion regime at the acceleration site follows Bohm diffusion. For a more realistic estimate, we assume that nuclei are subject to the same acceleration efficiency as electrons ($\chi \gtrsim 0.01$), resulting in an expected maximum energy of $Z$~PeV. For AX1A and AX1C the value is $E_{\rm max}^{AX1A/AX1C} \sim 0.2Z$~EeV, however, if the same acceleration efficiency of the electrons applies to the nuclei ($\chi \sim 2 \times 10^{-3}$), the maximum energy drops to $0.4 Z$~PeV. Given these estimates, it seems that protons and ions are some distance off reaching the UHECR regime, a finding which is in-keeping with other more general estimates in Cen A based on the current jet activity \citep[e.g.][]{massaglia_radio_2007,croston_2009,matthews_fornax_2018,matthews_ultrahigh_2019}. 

\section{Conclusions}
\label{sec:conclusions}
In this paper, we tested the hypothesis of knots as particle accelerators responsible for the VHE $\gamma$ radiation detected in Cen~A. The problem was treated in detail, combining multi-wavelength data with RHD simulations and particle acceleration theory. We focused mainly on the better-known AX1A, AX1C, AX2, and BX2 knots. We find that particle acceleration at jet-SW interaction sites can provide a plausible explanation for the X-ray and radio knots as well as the VHE $\gamma$-ray energy emission observed from the Cen A jet. The main conclusions developed here are listed below.

\begin{itemize}
    \item \textbf{Energetics}: In general, the model results are in good agreement with the radio and X-ray data for all the knots. Knots AX2 and BX2 should accelerate electrons at least up to $\sim 40$~TeV, but energies so high as $\sim 4$~PeV are expected. The inner AX1A and AX1C knots require a smaller acceleration efficiency, and the maximum energy achieved should not exceed $\sim 6$~TeV.
    
    The high values obtained for the maximum energy of electrons suggest the knots are the accelerators of the relativistic electrons demanded by the HESS measurements;%
    \item \textbf{Morphology}: The size obtained for knots AX1C, AX2, and BX2 agrees with the current estimations based on \textit{Chandra} X-ray data, indicating that the parameters selected for the SW are reasonable. This suggests that powerful WR stars as possible obstacles to generating these knots, which is plausible given the expectations~\citep{Muller_2023_conf} for the number of WR stars along the jet.
    
    The radius of the knot AX1A is more than twice that found by the X-ray measurements, which could indicate a weaker SW. The presence of a substructure in the AX1A knot challenges the explanation by an obstacle only, but interactions with AX1C could not be discarded.
    
    All models present an excess in the number of photons when compared to the X-ray images. We attribute this to the diffusion effect, not included in our model.

    \item $\gamma$\textbf{-rays}: Assuming the current estimations for the knot's magnetic field, we obtained that the emission of the knot centroid is not enough to account for the HESS measurements. The extended emission from electrons accelerated at the knots constitutes a satisfactory explanation for the VHE $\gamma$ radiation measured by HESS. As expected, the brighter knots in X-ray dominate the $\gamma$-rays band, especially AX1C and BX2. The $\gamma$-ray flux measured by Fermi-LAT could also be explained by the cooling, along the jet, of the electrons injected from the knots' shock. These electrons could be trapped in the lobes~\citep{fermi_inner_cenA}. Another possibility is a still lower maximum acceleration efficiency of the inner knots AX1A and AX1C.

    \item \textbf{Magnetic Fields:} Given the jump conditions obtained in this work, we propose that the $23~\mu$G magnetic field obtained by HESS using data from a region $2.4-3.6$~kpc away from the core corresponds to a magnetic field amplified by the knots of the F- and G-complexes.
    
    Assuming the validity of our model in the extended region of the knots, the sum of the $\gamma$-ray flux exceeds the measurements from HESS. By normalizing the $\gamma$-ray flux from BX2, we found that the magnetic field in the BX-complex should be closer to $\sim80~\mu$G, as estimated by \citet{Snios_2019}.
\end{itemize}

From the point of view of UHECR, we expect that the knots accelerate particles up to $\sim 0.2 Z$~EeV, in a highly efficient scenario. This is not enough to explain the more energetic $\sim 10^{19}$~eV particles. However, knots could inject heavier nuclei~\citep{wykes_2015,muller} from WR stars that will undergo subsequent acceleration in the jet and/or lobes.

Besides not being able to disentangle the emission from individual knots, the brighter AX1A/AX1C and BX2 have different maximum energies. In principle, measurements with high energetic and spatial resolution may indicate the dominant knots in the VHE band and test the acceleration efficiency in the knots. This constitutes the ideal scenario for tests using the CTAO~\citep{cta_book}, as is visible from Figure~\ref{fig:sed:sum_diffuse}. In the energy range $10 - 100$~TeV, the angular resolution of the CTAO will be $\sim0.02 - 0.03^\circ \approx 70 - 100'' \sim 1.3 - 1.8$~kpc~\citep{cta_book}. This angular resolution is enough to separate the contribution of the inner knots complexes (A, B, C, and E) from the outer ones (F and G) and the lobes since AX2 is located at $\sim20''$ and the F complex is at $\sim150''$ from the core. In a future publication, we will explore the potential of CTAO to study particle acceleration in Cen~A.

\section*{Acknowledgements}

CdO thanks Emma Elley for useful discussions. The authors acknowledge the National Laboratory for Scientiﬁc Computing (LNCC/MCTI, Brazil) for providing HPC resources of the SDumont supercomputer (http://sdumont.lncc.br) and the University of Oxford Advanced Research Computing (ARC) facility (\url{http://dx.doi.org/10.5281/zenodo.22558}).
This study was financed, in part, by the São Paulo Research Foundation (FAPESP), Brasil. Process Number 2023/16753-0, 	2021/01089-1, 2020/15453-4 and 2019/10151-2. JHM acknowledges funding from a Royal Society University Research Fellowship (URF\textbackslash R1\textbackslash221062).
This work made use of Astropy:\footnote{http://www.astropy.org} a community-developed core Python package and an ecosystem of tools and resources for astronomy \citep{astropy:2013, astropy:2018, astropy:2022}. We also gratefully acknowledge the use of the following software packages: visit~\citep{childs2005proceedings}, matplotlib~\citep{Hunter:2007}, pluto 4.4~\citep{Mignone_2007}, and Naima~\citep{naima}. This research has made use of the NASA/IPAC Extragalactic Database (NED) which is operated by the Jet Propulsion Laboratory, California Institute of Technology, under contract with the National Aeronautics and Space Administration.
This research has made use of the VizieR catalog access tool, CDS, Strasbourg, France (DOI : 10.26093/cds/vizier). The original description of the VizieR service was published in 2000, A\&AS 143, 23.
The scientific results reported in this article are based in part on observations made by the Chandra X-ray Observatory and published previously in cited articles.

\section*{DATA AVAILABILITY}
The data produced as part of this work are available from the authors on reasonable request.

\bibliographystyle{mnras}
\bibliography{main.bib}

%\begin{figure}
%    \centering
%    \includegraphics[width=\linewidth]{figs/Urad_zoom.pdf}
%    \caption{Jet shock for each knot with parabolic fit. The profile of AX1 and AX2 almost overlap.}
%    \label{fig:Urad_knots}
%\end{figure}

%\begin{figure}
%    \centering
%    \includegraphics[width=\linewidth]{figs/Spectrum.pdf}
%    \caption{Radiation field energy by unit of frequency in the shock position of the knots. The shock is approximated at $z_{\rm knot} - 0.1 R_{jet}$. \textcolor{red}{NEDD DUST CORRECTION}}
%    \label{fig:Urad_spec}
%\end{figure}

%\begin{figure}
%    \centering
%    \includegraphics[width=\linewidth]{figs/Emax_BX2_comp.pdf}
%    \caption{Break energy for the knot BX2 with $\chi=1$ and $\beta_m=800$ compared to $\chi=0.01$ (left), $\beta_m = 210$ (center), and AX2 (right).}
%    \label{fig:Ebreak_BX2}
%\end{figure}

%\begin{figure}
%    \centering
%    \includegraphics[width=\linewidth]{figs/comparison_knots.png}
%    \caption{Synchrotron SED predicted for the knots in different scenarios. $B=Bp/ \sqrt{\beta}$, $E_{\rm max} = \chi E_{\rm max}(\chi=1)$. Radius uses half of the knot size determined by Tanada as the distance to the SW position to determine the knot area. Diameter uses the knot size determined by Tanada. Resolution uses half the \textit{Chandra} angular resolution.}
%    \label{fig:sed:area}
%\end{figure}

%\begin{figure}
%    \centering
%    \includegraphics[width=\linewidth]{figs/test_bmin.png}
%    \caption{B value predicted for BX2.}
%    \label{fig:sed:ax2_bx2}
%\end{figure}

\appendix
\section{Radiation fields for non-thermal emission} \label{sec:app:radiation_fields}

\begin{figure*}
    \centering
    \includegraphics[width=\linewidth]{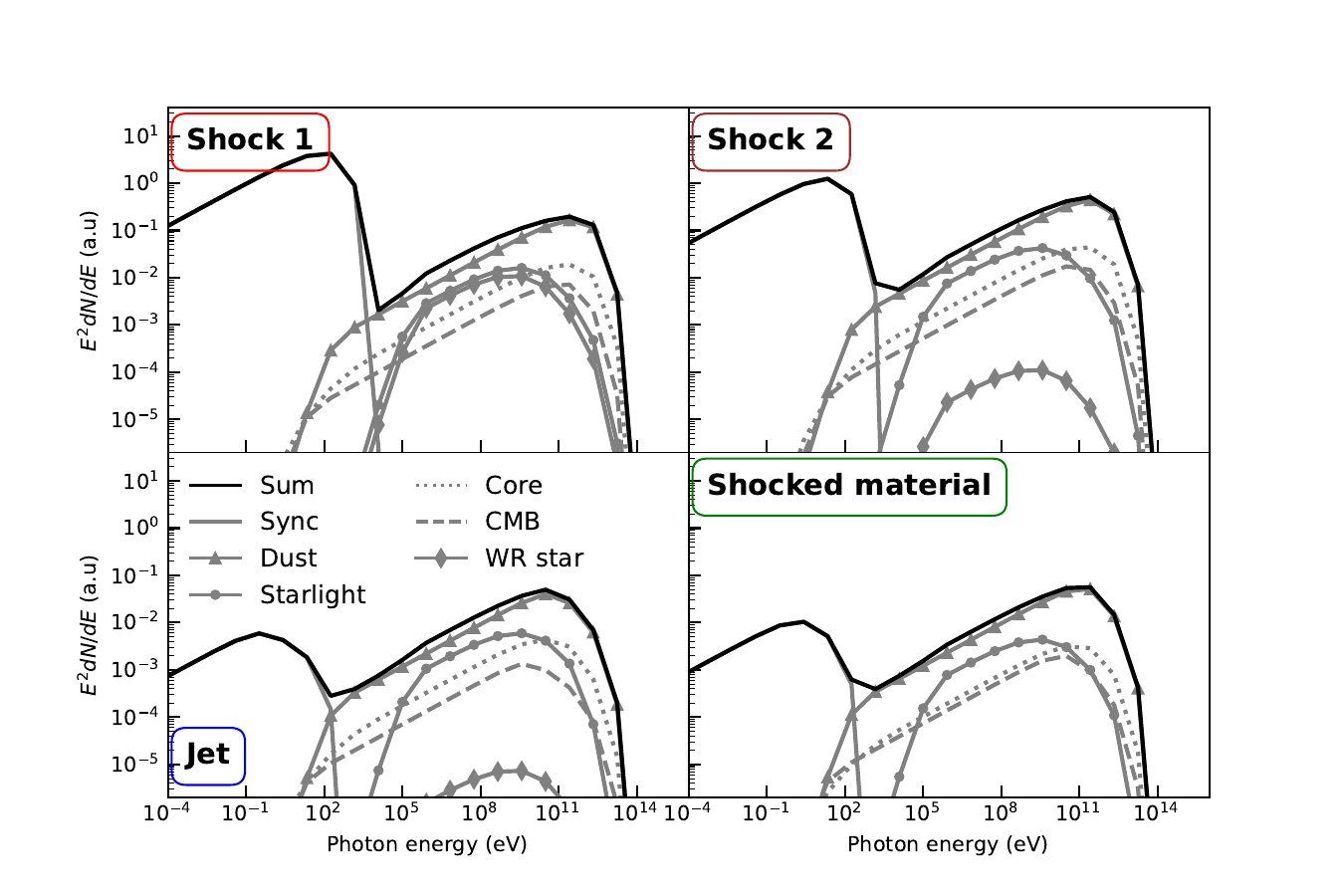}
    \caption{SED in different locations along AX1 knot. The positions are the same as shown in Figure~\ref{fig:sites:AX1_by_site}: shock 1 (red square), shock 2 (brown square), jet (blue square), and shocked material (green square).}
    \label{fig:sites:AX1_SED}
\end{figure*}

Figure~\ref{fig:sites:AX1_SED} shows the radiation emitted in each point of figure~\ref{fig:sites:AX1_by_site} via synchrotron and IC for all the radiation fields considered and the positions shown in Figure~\ref{fig:sites:AX1_by_site}. Here we show the profile for the inner knot AX1A, where the radiation fields are more intense, for $\chi=2\times10^{-3}$. Dust is the dominant radiation to generate VHE $\gamma$ rays in all cases. External Compton with starlight radiation significantly generates $\gamma$ rays in MeV/GeV scale. The synchrotron self-Compton contribution is negligible and is not shown.

%----------Apendix Figure ------------

\end{document}